\documentclass[fleqn,usenatbib]{mnras}

\usepackage{newtxtext,newtxmath}
\usepackage[normalem]{ulem}

\usepackage[T1]{fontenc}
\usepackage{ae,aecompl}

\usepackage{graphicx}	
\usepackage{amsmath}	
\usepackage{amssymb}	
\usepackage[colorinlistoftodos]{todonotes}

\title[A Blind Test of Reverberation Mapping Methods]{Do Reverberation Mapping Analyses Provide an Accurate Picture of the Broad Line Region?}

\author[S. W. Mangham et al.]
{S. W. Mangham,$^{1}$\thanks{E-mail: s.w.mangham@soton.ac.uk},
C. Knigge,$^{1}$,
P. Williams, $^{3}$
Keith Horne,$^{2}$
A. Pancoast, $^{3}$\newauthor
J. H. Matthews,$^{4}$
K. S. Long,$^{5,6}$
S. A. Sim,$^{7}$
and N. Higginbottom$^{1}$
\\
$^{1}$ Department of Physics and Astronomy, University of Southampton, Southampton, SO17 1BJ, UK\\
$^{2}$ SUPA School of Physics \& Astronomy, University of St. Andrews, North Haugh, St. Andrews KY16 9SS, UK\\
$^{3}$ Harvard-Smithsonian Center for Astrophysics, Garden Street, Cambridge, MA 02138, USA\\
$^{4}$ University of Oxford, Astrophysics, Keble Road, Oxford, OX1 3RH, UK\\
$^{5}$ Space Telescope Science Institute, 3700 San Martin Drive, Baltimore, MD 21218, USA\\
$^{6}$ Eureka Scientific, Inc., 2452 Delmer St., Suite 100, Oakland, CA 94602-3017, USA\\
$^{7}$ School of Mathematics and Physics, Queen's University Belfast, University Road, Belfast, BT7 1NN, UK}

\date{Accepted XXX. Received YYY; in original form ZZZ}

\pubyear{2015}

\hypersetup{draft}

\begin{document}
\label{firstpage}
\pagerange{\pageref{firstpage}--\pageref{lastpage}}
\maketitle

\newcommand{\cqsocombrmax}{$nan$}
\newcommand{\cqsocombrmean}{$11.7_{-2.5}^{+4.0}$}
\newcommand{\cqsocombrmedian}{$7.5_{-1.7}^{+2.4}$}
\newcommand{\cqsocombrmin}{$2.1_{-1.1}^{+1.2}$}
\newcommand{\cqsocombsigmar}{$12.3_{-2.7}^{+6.9}$}
\newcommand{\cqsocombtaumean}{$11.6_{-2.3}^{+3.1}$}
\newcommand{\cqsocombtaumedian}{$6.6_{-1.5}^{+1.9}$}
\newcommand{\cqsocombbeta}{$1.28_{-0.16}^{+0.20}$}
\newcommand{\cqsocombthetao}{$32_{-12}^{+36}$}
\newcommand{\cqsocombthetai}{$38_{-11}^{+18}$}
\newcommand{\cqsocombkappa}{$-0.35_{-0.08}^{+0.24}$}
\newcommand{\cqsocombgamma}{$>3.4$}
\newcommand{\cqsocombxi}{$0.29_{-0.19}^{+0.21}$}
\newcommand{\cqsocomblogmbh}{$8.20_{-0.17}^{+0.23}$}
\newcommand{\cqsocombfellip}{$0.33_{-0.21}^{+0.21}$}
\newcommand{\cqsocombfflow}{$0.64_{-0.34}^{+0.24}$}
\newcommand{\cqsocombthetae}{$36_{-23}^{+28}$}
\newcommand{\cqsocombsigmaturb}{$0.052_{-0.031}^{+0.025}$}
\newcommand{\cqsocombinflowoutflow}{$0.40_{-0.74}^{+0.23}$}
\newcommand{\cqsocomblogfrmssigma}{$nan_{-nan}^{+nan}$}
\newcommand{\cqsocomblogfmeansigma}{$nan_{-nan}^{+nan}$}
\newcommand{\cqsocomblogfmeanfwhm}{$nan_{-nan}^{+nan}$}
\newcommand{\cqsocombloverledd}{$nan_{-nan}^{+nan}$}

\newcommand{\cseyrmax}{$nan$}
\newcommand{\cseyrmean}{$16.9_{-1.0}^{+1.1}$}
\newcommand{\cseyrmedian}{$6.9_{-1.1}^{+1.3}$}
\newcommand{\cseyrmin}{$1.33_{-0.27}^{+0.27}$}
\newcommand{\cseysigmar}{$31.2_{-2.2}^{+1.4}$}
\newcommand{\cseytaumean}{$15.9_{-4.6}^{+5.5}$}
\newcommand{\cseytaumedian}{$4.1_{-1.5}^{+2.7}$}
\newcommand{\cseybeta}{$1.861_{-0.061}^{+0.077}$}
\newcommand{\cseythetao}{$<25$}
\newcommand{\cseythetai}{$>68$}
\newcommand{\cseykappa}{$0.06_{-0.41}^{+0.31}$}
\newcommand{\cseygamma}{$3.1_{-1.4}^{+1.3}$}
\newcommand{\cseyxi}{$0.47_{-0.30}^{+0.36}$}
\newcommand{\cseylogmbh}{$7.84_{-0.07}^{+0.08}$}
\newcommand{\cseyfellip}{$<0.23$}
\newcommand{\cseyfflow}{$0.47_{-0.29}^{+0.35}$}
\newcommand{\cseythetae}{$76_{-32}^{+7}$}
\newcommand{\cseysigmaturb}{$0.054_{-0.046}^{+0.025}$}
\newcommand{\cseyinflowoutflow}{$-0.07_{-0.18}^{+0.32}$}
\newcommand{\cseylogfrmssigma}{$nan_{-nan}^{+nan}$}
\newcommand{\cseylogfmeansigma}{$nan_{-nan}^{+nan}$}
\newcommand{\cseylogfmeanfwhm}{$nan_{-nan}^{+nan}$}
\newcommand{\cseyloverledd}{$nan_{-nan}^{+nan}$}

\newcommand{\ps}{{\sc PrepSpec}}
\newcommand{\memecho}{{\sc MEMEcho}}
\newcommand{\memsys}{{\sc MEMSYS}}

\newcommand{\lcofifteen}{{\sc LCOGT 2015}}
\newcommand{\lampeight}{{\sc LAMP 2008}}
\newcommand{\lampeleven}{{\sc LAMP 2011}}

\newcommand{\ngc}{Arp~151}

\newcommand{\et}{{et~al.\, }}

\newcommand{\mbh}{\mbox{M$_{\rm BH}$}}
\newcommand{\msun}{\mbox{M$_\odot$}}
\newcommand{\kms}{\mbox{\rm km~s$^{-1}$}}

\newcommand{\lya}{\mbox{\rm Ly$\alpha$}}
\newcommand{\ha}{\mbox{\rm H$\alpha$}}
\newcommand{\hb}{\mbox{\rm H$\beta$}}
\newcommand{\hg}{\mbox{\rm H$\gamma$}}
\newcommand{\civ}{\mbox{\rm C{\sc ~IV}}}
\newcommand{\siiv}{\mbox{\rm Si{\sc ~IV}}}
\newcommand{\heii}{\mbox{\rm He{\sc ~II}}}
\newcommand{\nv}{\mbox{\rm N{\sc ~V}}}

\newcommand{\dd}{\mbox{\rm d}}
\newcommand{\fracd}[2]{\frac{\displaystyle{#1}}{\displaystyle{#2}}}


\begin{abstract}
    Reverberation mapping (RM) is a powerful approach for determining the nature of the broad-line region (BLR) in active galactic nuclei. However, inferring physical BLR properties from an observed spectroscopic time series is a difficult inverse problem. Here, we present a blind test of two widely used RM methods: \memecho\ (developed by Horne) and CARAMEL (developed by Pancoast and collaborators). The test data are simulated spectroscopic time series that track the  H$\alpha$ emission line response to an empirical continuum light curve. The underlying BLR model is a rotating, biconical accretion disc wind, and the synthetic spectra are generated via self-consistent ionization and radiative transfer simulations. We generate two mock data sets, representing Seyfert galaxies and QSOs. 
The Seyfert model produces a largely {\em negative} response, which neither method can recover. However, both fail ``gracefully'', neither generating spurious results. For the QSO model both CARAMEL and expert interpretation of \memecho\'s output both capture the broadly annular, rotation-dominated nature of the line-forming region, though \memecho\ analysis overestimates its size by 50\%, but CARAMEL is unable to distinguish between additional inflow and outflow components. Despite fitting individual spectra well, the CARAMEL velocity-delay maps and RMS line profiles are strongly inconsistent with the input data. Finally, since the H$\alpha$ line-forming region is rotation dominated, neither method recovers the disc wind nature of the underlying BLR model. Thus considerable care is required when interpreting the results of RM analyses in terms of physical models.
\end{abstract}




\section{Introduction}
\label{sec:introduction}

Reverberation mapping (RM) has become a powerful tool for determining the physical properties of active galactic nuclei (AGN; \citealt{Peterson1993}). As the continuum of an AGN varies, each broad emission line (BEL) in its spectrum usually responds with a mean lag, $\tau$, on the order of hours to months, depending on the luminosity of the system and the specific transition involved \citep{Onken2004,Kaspi1999, Kaspi2005}. By associating this lag with the light travel time from the central engine to the line-forming region, one  obtains an estimate for the size of the broad line region (BLR), $R_{\rm BLR} \simeq c \tau$. Moreover, BELs in Type I AGN are substantially Doppler-broadened, so the width of a line is a measure of the velocity field in the BLR. If this is assumed to be roughly virialized -- $v_{\rm BLR} \simeq \sqrt{G M_{\rm BH}/R_{\rm BLR}}$ -- a black hole mass estimate can also be found immediately as $M_{\rm BH} \simeq c\tau_{\rm BLR} v_{\rm BLR}^2  / G$.

\begin{figure}
	\includegraphics[width=\columnwidth]{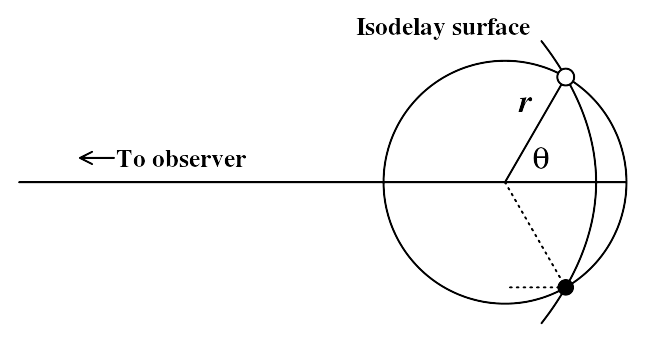}
    \label{fig:intro:isodelay}
    \caption{Isodelay contour from \protect{\citet{Peterson2004}}.}
\end{figure}

In reality, line-emitting material at any given distance from the central engine can produce a wide range of observed lags. Equivalently, any given lag can in principle be produced by line-emitting material across a wide range of distances. More specifically, isodelay surfaces around a point source form a set of concentric paraboloids of revolution centred on the line connecting the observer and the source (Figure~\ref{fig:intro:isodelay}; \cite{Peterson2004}). Which parts of an isodelay surface {\em actually} produce the observed line emission at a given lag -- and which isodelay surfaces actually contribute to a given line -- depends on the geometry, density and ionization state of the BLR, as well as on the inclination of the observer with respect to the system. In addition, different {\em parts} of a given line -- e.g. red and blue line wings -- will exhibit different responses, depending on the kinematics of the BLR. For example, if the BLR is  dominated by outflow kinematics, material moving towards the observer is also physically closest to the observer. The response of the blue wing is then expected to lead that of the red wing (see Figure~\ref{fig:background:tf_example}).

The true response of a given line to continuum variations is therefore  a distribution function over (at least\footnote{In principle, the line response also depends on the (possibly variable) source spectral energy distribution (SED), on the amplitude and shape of the continuum variations and on any physical changes in the BLR structure over the course of the observing campaign.}) time delay and radial velocity. Moreover, the shape of this function encodes information about the structure of the BLR -- its geometry, kinematics and ionization state -- on otherwise unresolvably small physical scales \citep{Welsh1991}. Thus, provided we can decode them, time- and velocity-resolved response functions (aka ``velocity-delay maps") provide us with a powerful tool for determining the physical nature of the BLR.

Several recent observational campaigns have obtained time-resolved spectroscopic data sets designed to allow the recovery of such 2-D response functions \citep{DeRosa2015a, Du2014}. These efforts have produced a wide range of apparent kinematic signatures, from simple rotation to both inflows \citep{Ulrich1996, Grier2013, Bentz2008, Bentz2010, Gaskell1988, Koratkar1989} and outflows \citep{Denney2009, Du2016}. This  variety of recovered response functions suggests that either the BLR is a much more heterogeneous entity than envisaged by current unification models, or that we are not correctly recovering and/or interpreting response functions from observational data. The aim of this paper is to test the latter possibility.

It is reasonable to worry about our ability to infer the physical structure of the BLR from the available observational data. Fundamentally, the issue is that this type of inference represents an ill-posed inverse problem. In particular, even though any time-resolved spectroscopic data set of finite quality contains {\em some} information about the BLR geometry and kinematics, this provides no guarantee of uniqueness. Thus, in principle, many physically different BLR structures might all produce very similar observational signatures. In practice, external constraints (aka ``regularization") are usually used  to select a unique solution from the set of models that are consistent with the data. The challenge is establishing that the selected solution is the physically correct one.

An excellent way to validate the inversion methods used in RM is to test them in a controlled manner against a \emph{known} response function. We can do this by calculating the response function for a physically-motivated BLR model and using this to generate a simulated spectroscopic time series that is comparable in quality to those provided by current observing campaigns \cite[e.g.][]{DeRosa2015a}. We can then apply the RM methods to
this mock data set and check the results against ``ground truth" (which is known in this case).

Here, we implement this idea by building on the results of
\cite{Mangham2017}. There, we used a radiative transfer and ionisation code \citep{Long2002} to produce physically-motivated response functions for one candidate BLR geometry, a rotating biconical disc wind \citep{Shlosman1993}. We now use these response functions, together with a driving continuum derived from an actual UV data set \citep{Fausnaugh2016}, to generate mock spectroscopic observing campaigns.

The two specific model response functions we use in our tests are based on a high-luminosity QSO and low-luminosity AGN disc wind model, respectively. One of these satisfies standard assumptions about response function behaviour (i.e. linear positive response with increasing luminosity), the other one does not. Taken as a pair, they provide a sensitive test of RM methods in very different regimes.

The remainder of this paper is organised as follows. In Section~\ref{sec:background}, we outline the theory behind RM. In Section~\ref{sec:method}, we describe the radiative transfer and ionisation code we use to generate our simulated response functions and explain how these response functions are then used to generate the synthetic spectroscopic time series for our mock observing campaigns. The authors of the two specific RM methods we test -- \memecho\ and CARAMEL -- then describe their respective analysis techniques. In Section~\ref{sec:results}, we present, assess and discuss the results obtained by each method. Finally, in Section~\ref{sec:conclusions}, we summarise our main conclusions.

\section{Background}
\label{sec:background}

\subsection{Reverberation Mapping: Basic Principles}
\label{sec:background:basics}

Reverberation mapping is based on the premise that the BLR reprocesses the ionising continuum from the central source into line emission. Fluctuations in the continuum propagate outwards at the speed of light, causing changes in the line emission when they interact with material in the BLR at later times. The time-scale on which this material responds -- i.e. the time-scale on which ionization equilibrium is established in the BLR -- is thought to be short compared to both the continuum variability and light-travel time-scales. The contribution of BLR material at position ${\bf \underline{r}}$ to the total line luminosity observed at time $t$ then depends only on the continuum luminosity at the earlier time $t-\tau$. Here, $\tau$ is the additional time required for light emitted by the central engine to reach the observer via a path that includes ${\bf \underline{r}}$.

\subsection{1-D Response Functions: Delay Distributions}
\label{sec:backgrounsd:1drf}

If the line response is perfectly linear -- i.e. if each location in the BLR always produces a fixed number of line photons per continuum photon -- the total line luminosity $L$ at time $t$ is a function of the continuum $C$ as
\begin{equation}
	L(t) = \int_{0}^{\infty} C(t - \tau) \, \Psi(\tau) \, d\tau,
	\label{eqn:background:1d_rf}
\end{equation}
where $\Psi(\tau)$ is the so-called {\em 1-D transfer function} or {\em delay distribution}. More specifically, $\Psi(\tau)$ is the weighted average reprocessing efficiency of all parts of the BLR that contribute to the response observed at delay $\tau$.

In reality, line responses are {\em not} perfectly linear. A better approximation is to linearise the response around reference (e.g. long-term average) continuum and line luminosities $C_0$ and $L_0$ (e.g. Horne 1994):
\begin{equation}
	C(t) \simeq C_0 + \Delta C(t),
	\label{eqn:background:cont_lin}
\end{equation}
and
\begin{equation}
	L(t) \simeq L_0 + \Delta L(t).
    \label{eq:linearize}
\end{equation}
We can then define the \emph{response function} $\Psi_R$ to refer to only the variable parts of the luminosities,
\begin{equation}
	\Delta L(t) = \int_{0}^{\infty} \Delta C(t - \tau) \, \Psi_R(\tau) \, d\tau.
    \label{resp0}
\end{equation}
This model can deal with {\em globally} non-linear responses, so long as the continuum variations are small enough to ensure that the {\em local} response remains approximately linear.

Conceptually, the 1-D response function, $\Psi_R(\tau)$ describes how the line emission produced by a sharp continuum pulse is spread across a range of time delays, $\tau$. Mathematically, the response function is just the partial derivative
\begin{equation}
	\Psi_R(\tau) = \frac{\partial L(\tau)}{\partial C(t-\tau)}.
\label{resp1}
\end{equation}
We can make the dependence of the response function on the local conditions within the BLR more explicit by writing it in terms of the line luminosity {\em per unit volume},
\begin{equation}
	\Psi_R(\tau) = \int_{V_{\rm BLR}} \frac{\partial l({\bf\underline{r}})}
    {\partial C(t-\tau)} \,\, \delta(\tau({\bf\underline{r}})) \,\,dV.
	\label{resp2}
\end{equation}
Here, the integral on the right-hand-side is over the entire volume of the BLR, $\partial l({\bf\underline{r}}) / \partial C(t-\tau)$ is the line response {\em per unit volume} at location {\bf\underline{r}}, and the delta function ensures that only locations that produce the correct delay contribute to the total response at $\tau$.

In some previous works on reverberation mapping, there has been ambiguity over the use of the terms \emph{transfer} and \emph{response} functions as they refer to either the \emph{linear} or \emph{linearized} forms of the reverberation mapping equation. \citet{Mangham2017} shows that the two assumptions produce markedly different results, and clarity is important. Thus, in this work, we use the term \emph{transfer} function to refer to the parameter $\Psi$ from the linear form of the equation (Equation \ref{eqn:background:1d_rf}, also equivalent to the \emph{emissivity-weighted response function} $\Psi_{\rm EWRF}$ of \citet{Goad1993, Mangham2017}). We use the term \emph{response} function to refer to the parameter $\Psi_R$ from the linearized form of the equation (Equation \ref{resp0}).

\subsection{2-D Response Functions: Velocity-Delay Maps}
\label{sec:background:2drf}

The 1-D response function can be calculated straightforwardly from Equation~\ref{resp2} for any given BLR geometry and emissivity distribution. However, many different physical models of the BLR could, in principle, produce the same delay distribution.

One way to partially break this degeneracy is to add kinematic, i.e. velocity, information. We can do this by splitting our emission line light curve into distinct radial velocity bins and constructing the delay distribution separately for each bin. This immediately leads to velocity-resolved versions of Equations~\ref{resp0}-\ref{resp2}:
\begin{equation}
	\Delta L(v,t) = \int_{0}^{\infty} \Delta C(t - \tau) \, \Psi_R(v,\tau) \, d\tau,
\label{resp10}
\end{equation}
\begin{equation}
	\Psi_R(v,\tau) = \frac{\partial  L(v,\tau)}{\partial C(t-\tau)},
\label{resp11}
\end{equation}
\begin{equation}
	\Psi_R(v,\tau) = \int_{V_{\rm BLR}} \frac{\partial l({\bf\underline{r}})}
    {\partial C(t-\tau)}  \, \delta(\tau({\bf\underline{r}})) \, \delta (v({\bf\underline{r}})) \, dV.
\label{resp12}
\end{equation}
Here, $v({\bf\underline{r}})$ is the radial velocity at position ${\bf\underline{r}}$ in the BLR. The 2-D response function, $\Psi_R(v,\tau)$, is also often referred to as the {\em velocity-delay map} of the BLR.

If we make the assumption that within the limit of small changes around $C_0$ the dependence of the line luminosity on the driving continuum is a power law, i.e.
\begin{equation}
    L(v,\tau) = L_0(v,\tau) \bigg( \frac{C}{C_0} \bigg) ^\eta,
    \label{eqn:background:2drf:eta}
\end{equation}
then we can re-express the response function equation \ref{resp11} in terms of the dimensionless \emph{responsivity} $\eta$ as
\begin{equation}
    \Psi_R(v, \tau) = \eta \frac{L_0(v, \tau)}{C_0}.
\end{equation}
Broadly, then, responsivity is a measure of whether a change in the driving continuum results in an increase or \emph{decrease} in line emission, and can be described both for a region in the response function as $\eta(v, \tau)$, or for a point within the BLR as $\eta({\bf\underline{r}})$. Of note is that for a system where globally $\eta = 1$, the response function $\Psi_R$ becomes equivalent to the transfer function $\Psi$.

\subsection{Reverberation Mapping in Practice: ~~~~ ~~~ ~~~ ~~~ ~~~ Constructing a Data-Driven BLR Model}
\label{sec:background:inversion}

\begin{figure}
	\includegraphics[width=\columnwidth]{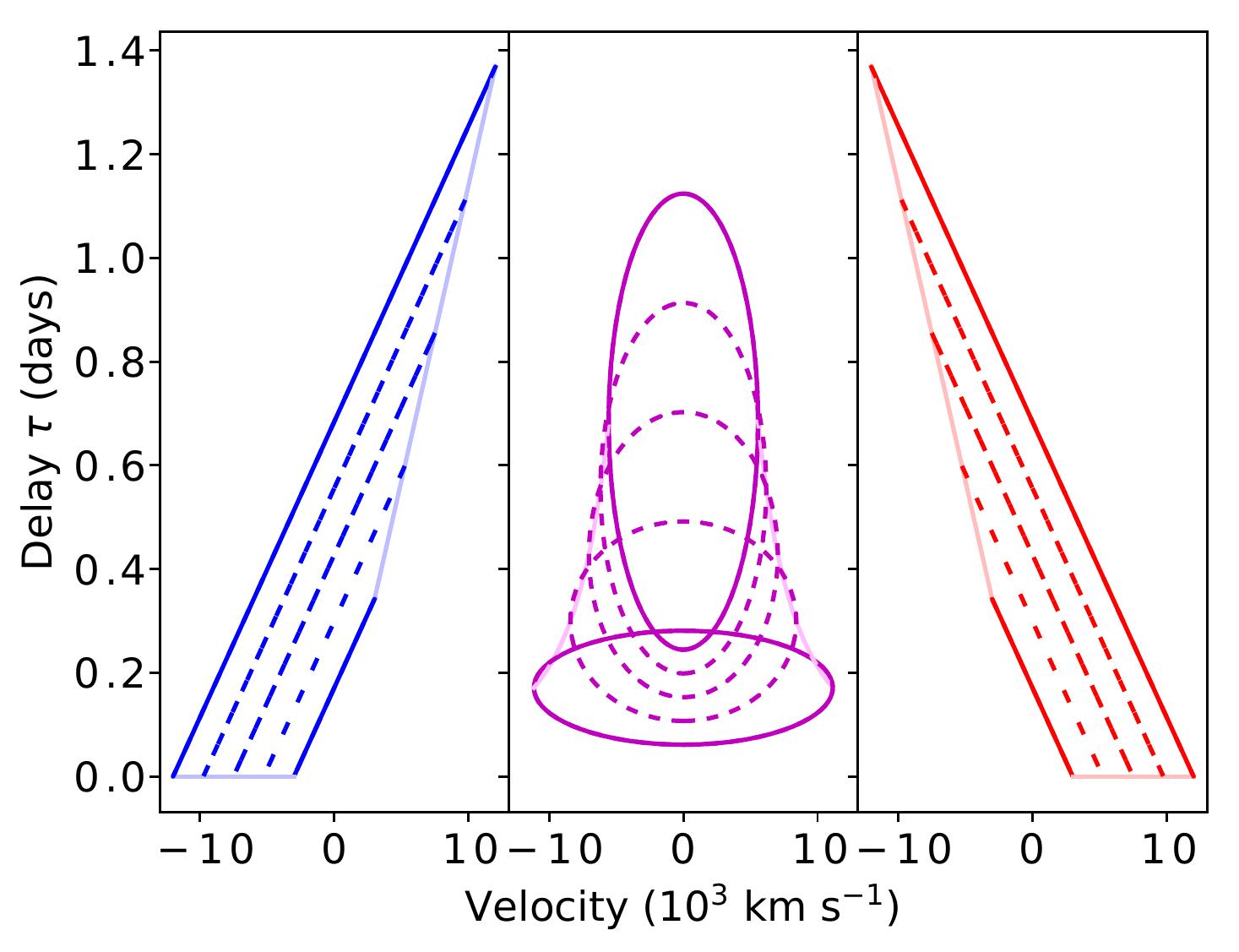}
	\begin{flushright}
    	\includegraphics[width=.9\columnwidth]{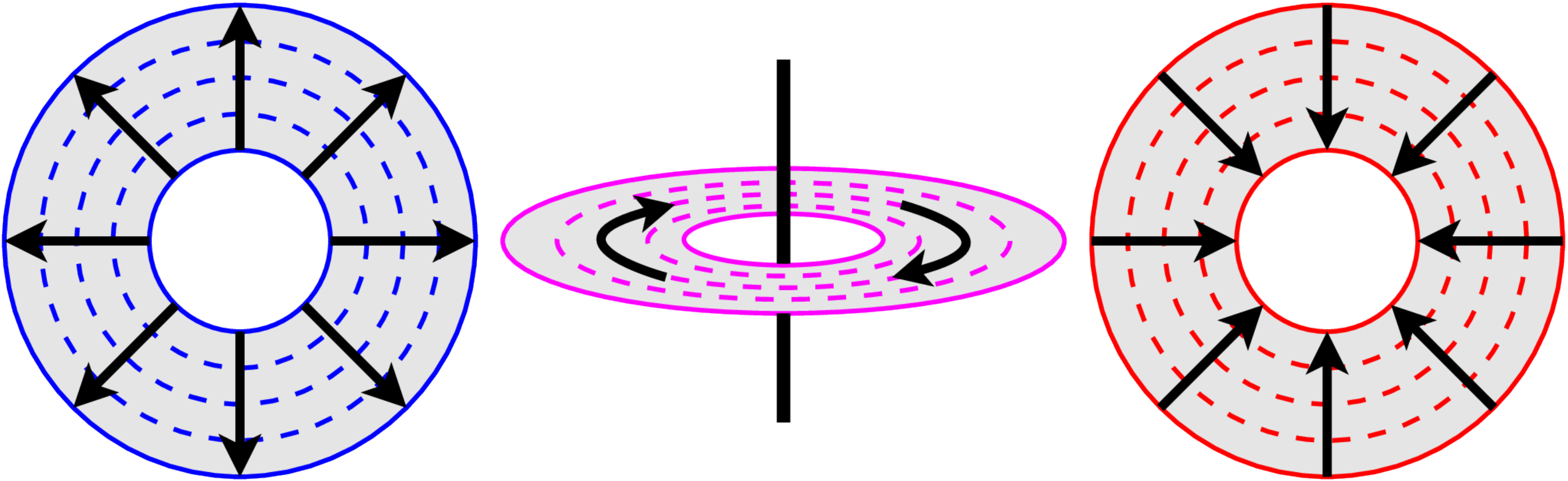}
    \end{flushright}
    \caption{Outline response functions and schematics for Hubble-type spherical outflow \textcolor{blue}{\textbf{(left)}}, a rotating Keplerian disc viewed at a $20^{\circ}$ angle \textcolor{magenta}{\textbf{(centre)}}, and Hubble-type spherical inflow \textcolor{red}{\textbf{(right)}}. Winds extend from $r_{\rm min}=20r_{g}$ to $r_{\rm max}=200r_{g}$ for an AGN of mass $10^{7} M_{\odot}$. Hubble out/inflows have $V(r_{\rm min})=\pm 3\times 10^{3}$ km s$^{-1}$. Solid lines denote the response from the inner and outer edges of the winds, dotted lines from evenly-spaced shells within the wind. Pale lines describe the edge of the velocity-delay shape of the response function.}
	\label{fig:background:tf_example}
\end{figure}

The addition of kinematic information lifts some of the degeneracy associated with $\Psi_R(\tau)$. For example, inflows and outflows that differ only in the sign of their velocity vector will produce identical 1-D response functions, but very different velocity-delay maps. As illustrated in Figure~\ref{fig:background:tf_example} (c.f. Horne 1994), inflow [outflow] kinematics are generally expected to produce a ``red-leads-blue" [``blue-leads-red"] signature in $\Psi_R(v, \tau)$. Similarly, pure rotation tends to produce symmetric velocity-delay maps, whose envelope is defined by the dependence of the rotation speed on distance from the central object.

However, in physically motivated models, such as the rotating disc wind model of \citet{Matthews2016}, we may expect to see a mix of these signatures. Moreover, the geometry of the line-emitting region will also depend strongly on the ionisation structure of the BLR. In fact, the line response may even be {\em negative} in parts of the velocity-delay space, as the associated sections of the wind become over-ionised and stop emitting. These effects can significantly complicate the interpretation of $\Psi_R(v, \tau)$ \citep{Mangham2017}.

Fundamentally, the issue is that even 2-D reverberation mapping is non-unique: a given data set, $C(t)$ and $L(v,t)$, may be consistent with many different physical BLR models. This is easy to understand. A full physical picture of the BLR requires us to specify at least the density and velocity at each position. Suppose we were to discretize the BLR spatially and kinematically into just 10 bins in each of these 7 parameters (position vector, velocity vector, density). The resulting model space then contains $10^7$ distinct parameter combinations, which vastly exceeds the number of data points in any realistic set of observations.  In order to find a unique solution for this type of ill-posed inverse problem, additional ``regularization" constraints have to be provided (e.g. based on physics, geometry, kinematics, smoothness, etc).

Two different approaches have so far been used to develop a data-driven physical picture of the BLR from time- and velocity-resolved observational reverberation mapping campaigns. In the first approach, the primary aim is the construction of $\Psi_R(v, \tau)$ from the data. Even this is an inverse problem, but determining a 2-D response function is a much more constrained problem than inferring an at least 7-D physical description of the BLR. The interpretation of the velocity-delay map in terms of an underlying physical picture is up to the user in this approach. Typically, this is based on a qualitative comparison of the recovered $\Psi_R(v, \tau)$ to the response functions produced by simple toy models, such as those in Figure~\ref{fig:background:tf_example}. We will refer to this as the ``inverse approach". In our tests, this approach is represented by \memecho\ \citep{Horne1994}, which is described more fully in Section~\ref{sec:method:memecho}.
The SOLA method \citep{Pijpers1994}, regularized linear inversion \citep{Krolik1995a, Skielboe2015} and the Sum-of-Gaussians method \citep{Li2016} all belong to this class as well.

The second approach is to build a highly flexible physical model of the BLR, whose parameters can be adjusted to fit the observed $L(v,t)$, using the observed $C(t)$. In principle, this approach does not require $\Psi_R(v, \tau)$ to be calculated explicitly for either the model or the data. In practice, the response function of the best-fitting model still provides a useful visualization tool and is easy to calculate from Equation~\ref{resp12}. The advantage of this approach is that the optimal fit parameters immediately provide a basic physical picture of the BLR. A key risk is that the parametrization of the model is incapable of describing the true BLR. This method also requires the implementation of all the physics needed to predict $L(v,t)$ from $C(t)$ for any given set of model parameters. We will refer to this as the ``forward-modelling approach". It is represented in our test by {\sc CARAMEL} \citep{Pancoast2011, Pancoast2014, Pancoast2014a}, which models the BLR as a population of reflecting, non-interacting clouds. CARAMEL is described more fully in Section~\ref{sec:method:caramel}.

\section{Methods}
\label{sec:method}

\subsection{Simulating an Observational Campaign}
\label{sec:method:fundamentals}

\subsubsection{Line Formation in a Rotating disc Wind}
\label{sec:method:fundamentals:line}
We use the radiative transfer and ionisation code \textsc{Python} to simulate the formation of the H$\alpha$ emission line in a rotating, biconical accretion disc wind model of the BLR. \textsc{Python} has already been described several times in the literature \citep{Long2002, Sim2005, Noebauer2010, Higginbottom2013, Higginbottom2014, Matthews2015, Matthews2016}, so we provide only a brief description of it here. Given a specification of the radiation sources, as well as of the disc wind geometry and kinematics, the code performs an iterative Monte Carlo ionization and radiative transfer simulation. It follows the paths of photons generated by a central X-ray source, an accretion disc and the wind itself through the system, records their interactions with the wind, and calculates their effect on the local temperature and ionization state. Once all photons in a given iteration have traversed the grid, the temperature and ionisation state of each wind cell is updated. This changes the emission profile and opacity of the wind, so the radiation transfer process is repeated. The ionisation state is then recalculated, and this process is iterated until the temperature and ionisation state of the wind have converged. The converged wind model is then used to generate detailed spectra for a range of user-specified observation angles.

\begin{figure}
	\includegraphics[width=\columnwidth]{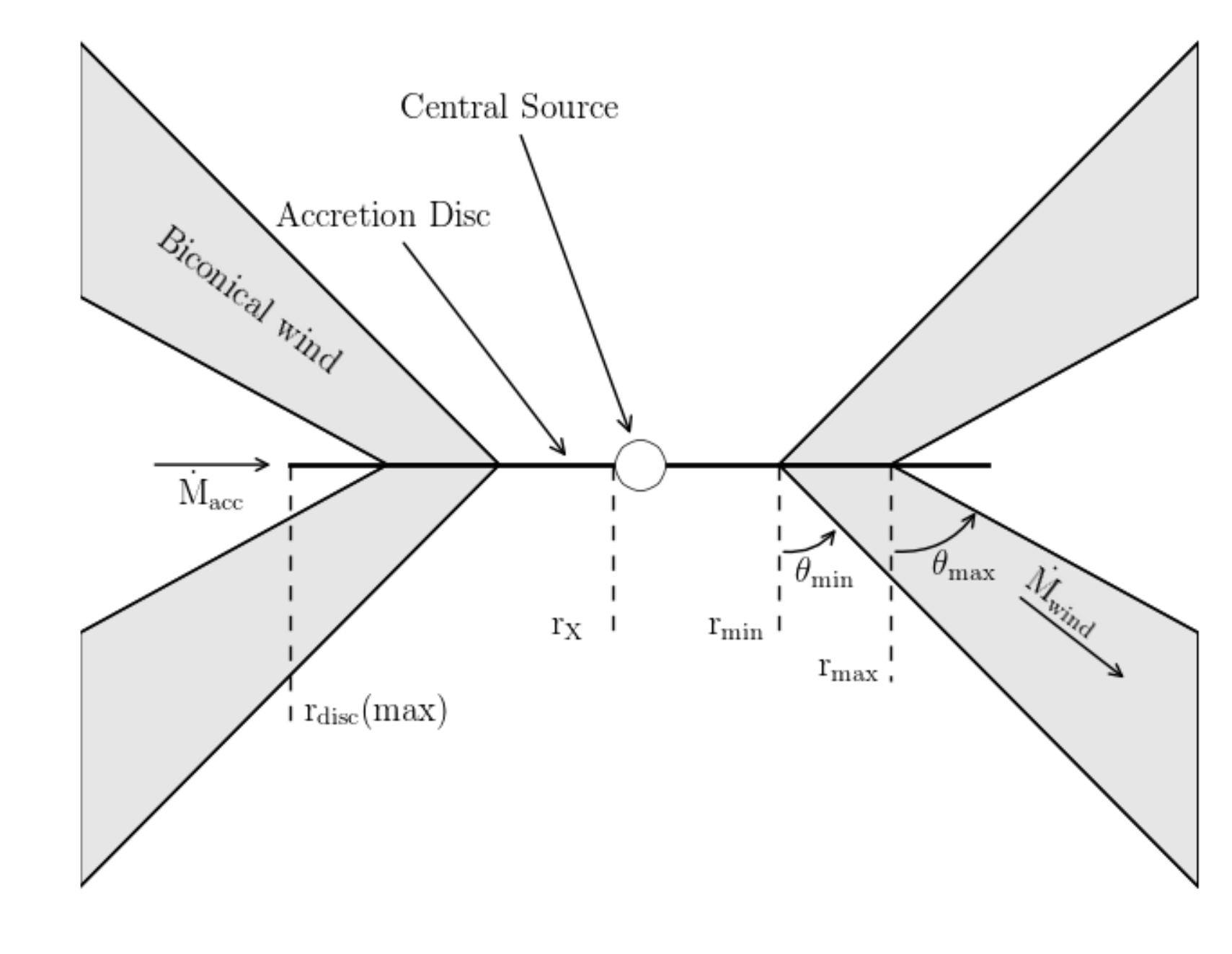}
    \caption{Sketch of the biconical disc wind geometry from \protect\cite{Matthews2016}.}
    \label{fig:background:geometry}
\end{figure}

\begin{figure}
	\includegraphics[width=\columnwidth]{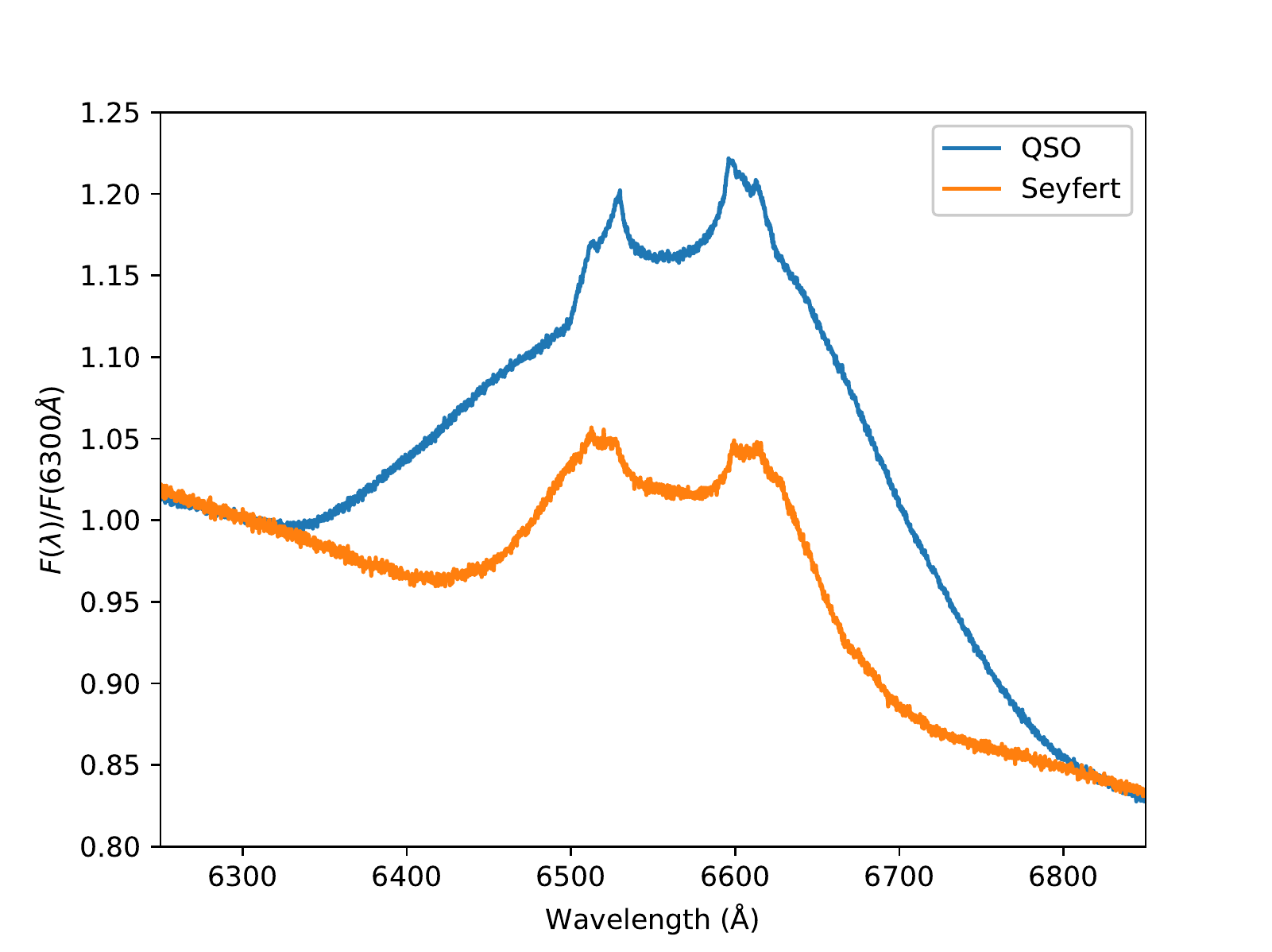}
	\caption{Mean line profiles for our the H$\alpha$ line in our QSO and Seyfert models.}
    \label{fig:method:profile}
\end{figure}

\begin{table*}
	\small
	\caption{Model parameters}
	\label{table:params}
	\begin{tabular}{l l | c c c c }
    	\hline \hline
		Parameter 				& Symbol, units					& Seyfert & Seyfert, rescaled 	& QSO & QSO, rescaled\\
		\hline
		Original SMBH mass 		& $M_{\rm BH}$, ${\rm M}_{\odot}$ 	& $10^{7}$	& $1.33\times 10^{8}$ & $10^{9}$ & $1.94\times 10^{8}$ \\
		Accretion rate			& $\dot{M}_{\rm acc}$, M$_{\odot}$ yr$^{-1}$		& 0.02 & 			& 5 &\\
								& $\dot{m}_{\rm acc}$, $\dot{\rm M}_{\rm Edd}$		& $\approx 0.1$	& 	& $\approx 0.2 $ &\\
		X-ray power-law	index	& $\alpha_{X}$					& $-0.9$ &						& $-0.9$ &\\
		X-ray luminosity		& $L_{X}$, erg s$^{-1}$	& $10^{43}$	& 					& $10^{45}$ & \\
        X-ray source radius		& $r_{X}$, r$_{g}$		& 6 & 							& 6	&\\
        						& $r_{X}$, cm			& $8.8 \times 10^{12}$	& $1.17 \times 10^{14}$		& $8.8 \times 10^{14}$	&
                                $1.71\times 10^{14}$\\
		Accretion disc radii	& $r_{\rm disc}$(min)			& $r_{X}$ 			& 			& $r_{X}$ &\\
								& $r_{\rm disc}$(max), r$_{g}$	& 3400 				& 			& 3400 &\\
								& $r_{\rm disc}$(max), cm		& $5 \times 10^{15}$ & $6.63\times 10^{16}$	& $5 \times 10^{17}$ & $9.69\times 10^{16}$\\
                                & $r_{\rm disc}$(max), ld		& $1.93$	& $25.6$	& $193$ 	& $37.41$\\
		Wind outflow rate 		& $\dot{M}_{w}$, M$_{\odot}$ yr$^{-1}$		& 0.02 & 	& 5 &\\
		Wind launch radii		& $r_{\rm min}$, r$_{g}$	& 300 & 				& 300 &\\
								& $r_{\rm min}$, cm	& 	$4.4 \times 10^{14}$ 	& $5.83\times 10^{15}$	& $4.4 \times 10^{16}$ & $8.53\times 10^{15}$\\
                                & $r_{\rm min}$, ld		& $0.1699$	& $2.251$	& $16.99$ 	& $3.293$\\
								& $r_{\rm max}$, r$_{g}$ 	& 600 & 			& 600 &\\
								& $r_{\rm max}$, cm		& $8.8 \times 10^{14}$ 	& $1.17 \times 10^{16}$	& $8.8 \times 10^{16}$ & $1.71\times 10^{16}$\\
								& $r_{\rm max}$, ld		& $0.3397$ 	& $4.517$	& $33.97$	& $6.602$\\
		Wind launch angles		& $\theta_{\rm min}$				& $70^{\circ}$ & 				& $70^{\circ}$ &\\
								& $\theta_{\rm max}$				& $82^{\circ}$ & 				& $82^{\circ}$ & \\
		Terminal velocity		& $v_{\infty}(r_0)$	 			& $v_{\rm esc}(r_0)$ & 				& $v_{\rm esc}(r_0)$	&\\
		Acceleration length		& $R_{\rm v}$, cm			& $10^{16}$ & $1.33\times 10^{17}$	& $10^{19}$ & $1.94\times 10^{18}$\\
		Acceleration index 		& $\alpha$						& 1.0 & 						& 0.5 &\\
		Filling factor 			& $f_V$							& 0.01 & 						& 0.01 &\\
		Viewing angle 			& $i$							& 40$^{\circ}$ & 				& 40$^{\circ}$ & \\
		Number of photons		& 								& $6\times 10^{9}$ & 			& $1\times 10^{10}$ & \\ \hline \hline

\end{tabular}
\end{table*}

The basic disc wind geometry we adopt for the BLR is illustrated in Figure~\ref{fig:background:geometry}. For the purpose of testing RM inversion methods, we consider two specific parameter sets: the Seyfert and QSO models from \citet{Mangham2017}. These were chosen because they represent plausible, physically-motivated response function signatures and display a range of behaviours that allow us to investigate the capabilities of inversion tools to cope with both straightforward and unusual response functions (see Section~\ref{sec:method:fundamentals:response}). All relevant parameters of these models -- along with brief explanations of what they encode -- are provided in Table~\ref{table:params}. The corresponding mean line profiles are shown in Figure~\ref{fig:method:profile}.

\subsubsection{Creating Response Functions}
\label{sec:method:fundamentals:response}

We use \textsc{Python} to generate 2-D response functions using the methodology described in \citet{Mangham2017}. Briefly, the response function, $\Psi_R(v,\tau)$, describes how a change in line emission at time $t$ depends upon changes in the continuum across a range of previous times $t-\tau$. This requires making the assumption that the response function itself is \emph{not} dependent on the incident continuum luminosity. If this is the case, then the response function can also be mirrored in time: it describes not only how the line emission observed at time $t$ depends on the continuum at time $t-\tau$, but \emph{also} how the continuum at time $t$ propagates out to change the line emission at time $t+\tau$. As a result of this, we can derive the response function $\Psi_R$ by forward-modelling. An instantaneous change in continuum luminosity at time $t$ drives changes in line emission at a range of times $t+\tau$. These changes can be calculated by tracking the photons in our Monte Carlo simulation and calculating their arrival time delay, $\tau$, from their path. These photons are then binned by velocity and delay to produce a response function $\Psi_{R}(v, \tau)$.

In order to match the average $\approx 3$ day delays seen in H$\beta$ for NGC~5548 \citep{Bentz2013}, we re-scale the time axis in our response functions so that the peak delay occurs at $\approx 3$ days (the exact peak delays in our rescaled models are 2.25 days for the Seyfert model and 4.05 days for the QSO model). The resulting rescaled response functions no longer correspond directly to a fully self-consistent BLR model, but this is not a concern for the purpose of testing RM inversion methods. Here, our main requirement is simply to have known response functions that (a) are characterised by an empirically-motivated mean delay, and (b) account for the complexities associated with physically motivated BLR models (e.g. mixed kinematics, complex emissivity distributions, negative responses).

\begin{figure*}
  \begin{minipage}{\columnwidth}
      \includegraphics[width=\columnwidth]{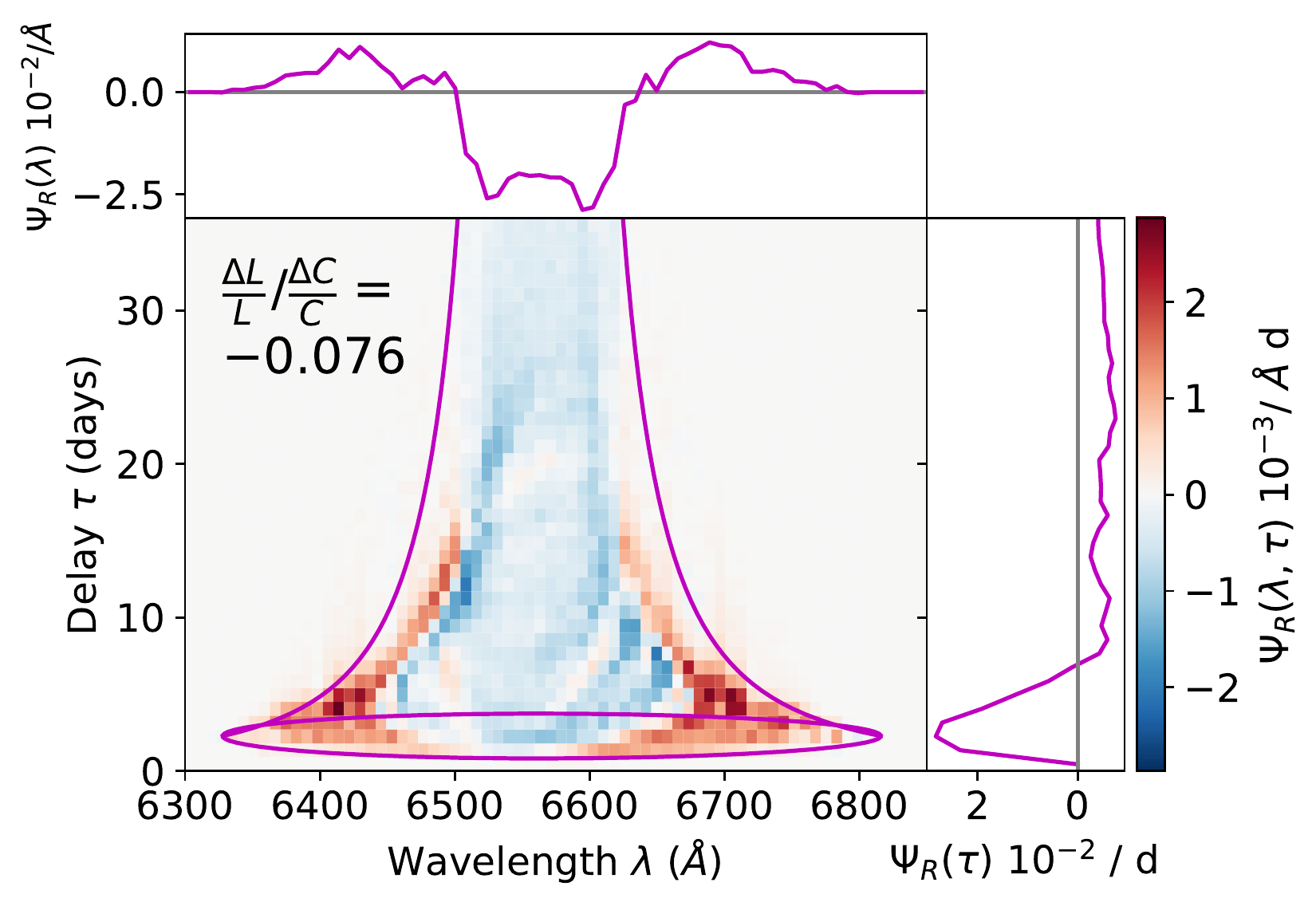}
      \caption{The `true' velocity-resolved response function (lower) for H$\alpha$ in
          the Seyfert model, rescaled to a peak delay of $\approx 3$ days.}
      \label{fig:method:response_sey}
  \end{minipage}\hfill\begin{minipage}{\columnwidth}
	\includegraphics[width=\columnwidth]{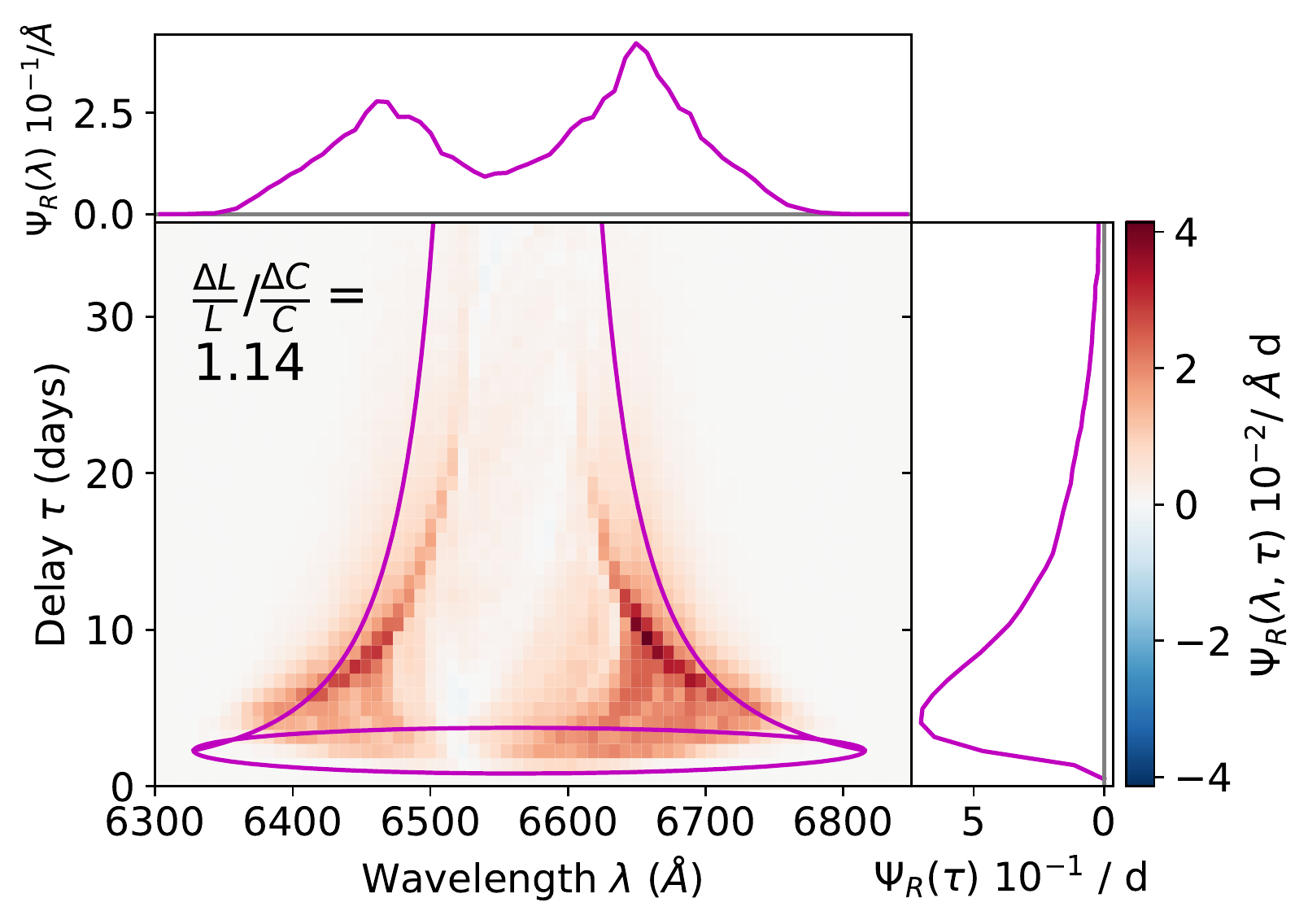}
    \caption{The `true' velocity-resolved response function (lower) for H$\alpha$ in
        the QSO model, rescaled to a peak delay of $\approx 3$ days.}
    \label{fig:method:response_qso}
\end{minipage}
\end{figure*}

The response functions calculated for our two models in this way are shown in Figures~\ref{fig:method:response_sey} and \ref{fig:method:response_qso}. A full discussion of these velocity-delay maps has already been provided by \citet{Mangham2017}, so here we only highlight some salient features. In both Seyfert and QSO models, the line emission and response is dominated by the dense, rotation-dominated base of the disc wind (c.f. Figures~\ref{fig:geometry-emissivity} and ~\ref{fig:geometry-responsivity} below). Correspondingly, the mean wavelength-dependent response is double-peaked in both cases. Both response functions also clearly show the virial envelope associated with this part of the outflow. The emissivity distributions appear double-lobed, rather than quadruple-lobed as in \citet{Murray1996}

A key difference between the models is that the H$\alpha$ response is always positive in the QSO model, but not in the Seyfert model. In fact, at low Doppler shifts and long delays, the velocity-delay map for the Seyfert model is so dominated by {\em negative} responsivities that the \emph{net} line response is also negative. As discussed in \citet{Mangham2017}, the Seyfert model includes substantial emission from the extended, low-density parts of the wind. As the ionising continuum luminosity increases, these regions become over-ionised, which reduces the H$\alpha$ emission produced at large radii. By contrast, the denser wind base is not over-ionised and responds positively to the increased continuum luminosity.

The combination of a net negative response and a change in the characteristic emission radius with ionising continuum is physically plausible. It has been observed in NGC~5548, for example, \citet{Cackett2006}, although there the BLR radius appears to \emph{increase} with increasing ionising luminosity. A possible negative correlation of the H$\alpha$ response with ionising continuum may also be present in the AGN STORM dataset of \citet{Pei2017} (see epoch T2 in their Figure~7).

\subsubsection{Creating Time-Series of Spectra}
\label{sec:method:tss}

\begin{figure}
	\includegraphics[width=\columnwidth]{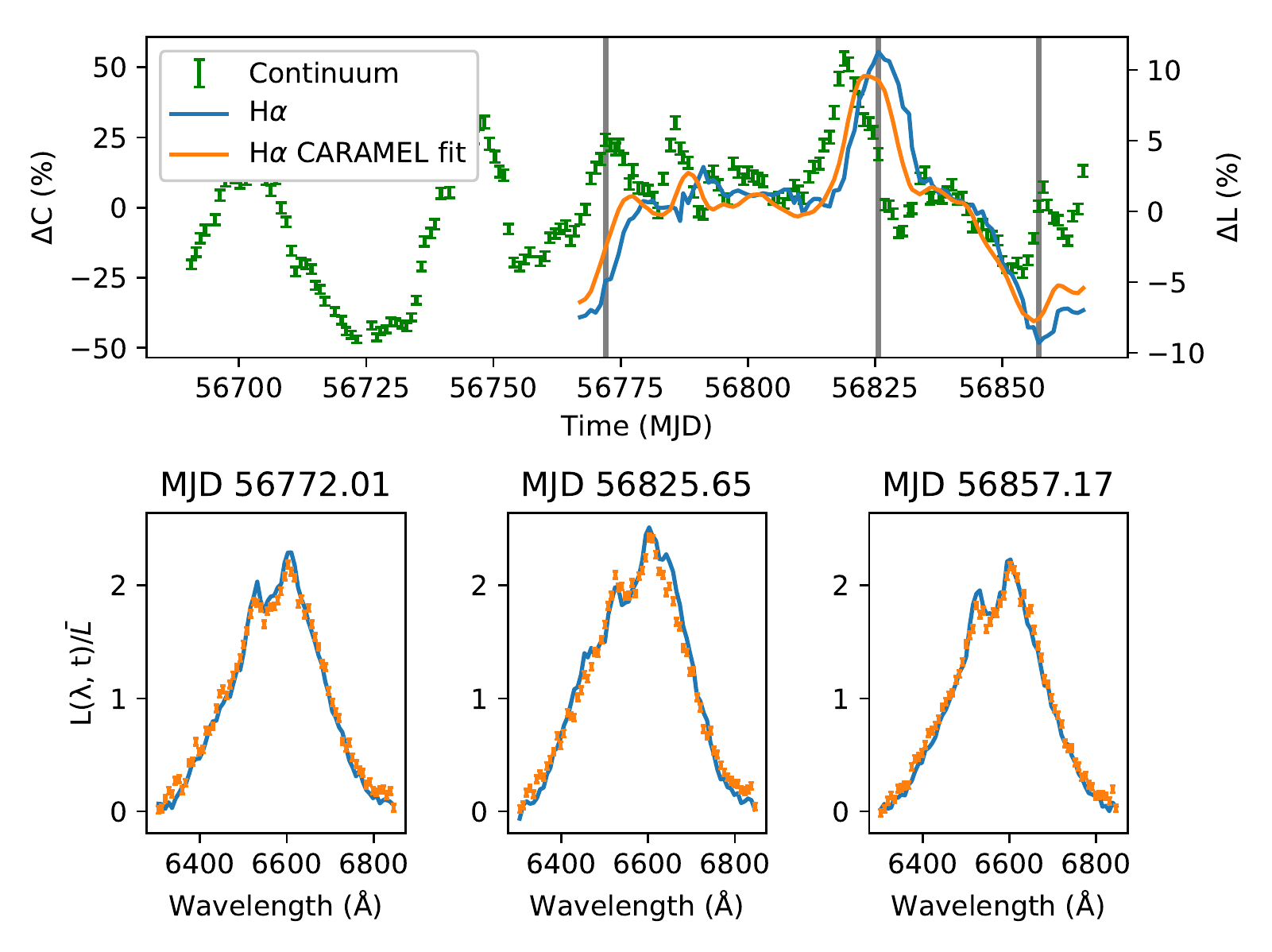}
	\caption{Driving continuum and line light-curves for our model and the CARAMEL fit (upper). Spectra associated with three `observational' times (indicated by grey vertical lines on the light-curves), and their CARAMEL fit (lower).}
    \label{fig:results:lightcurve_spectra}
\end{figure}

\begin{figure*}
	\begin{minipage}{.48\textwidth}
		\includegraphics[width=\columnwidth]{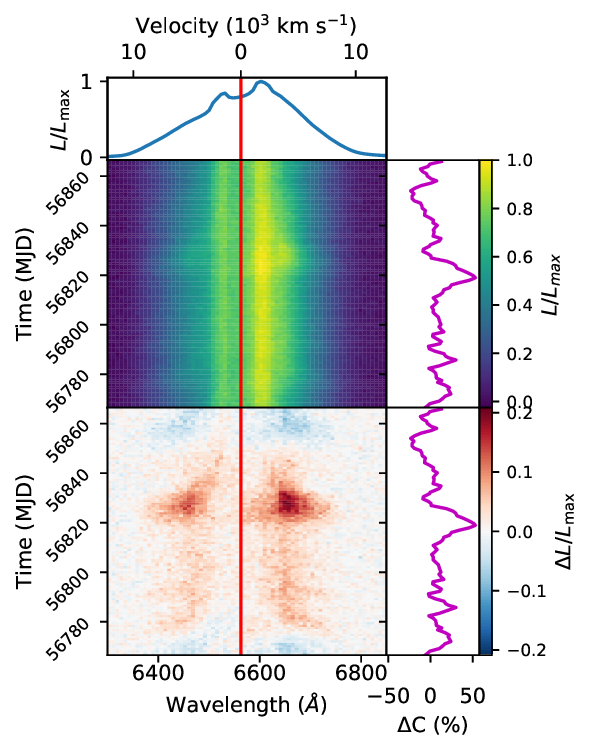}
    \end{minipage}\hfill
    \begin{minipage}{.48\textwidth}
    	\includegraphics[width=\columnwidth]{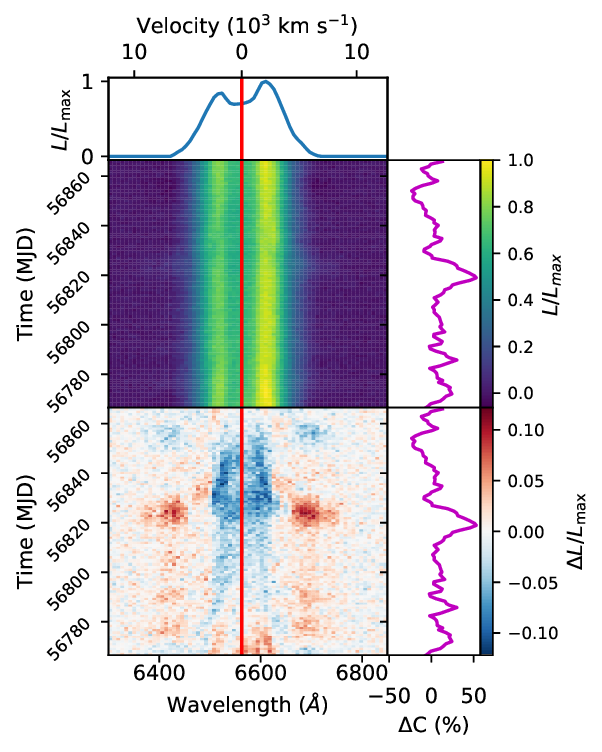}
    \end{minipage}

    \caption{Trailed spectrograms generated for the continuum-subtracted H$\alpha$ lines of our QSO (left) and Seyfert (right) models over a simulated observing campaign of 98.9 days.}
    \label{fig:method:tss}
\end{figure*}

Actual RM campaigns do not observe response functions; they observe spectroscopic time series. In order to test RM methods, we therefore have to generate such time series from our models. Equation \ref{resp10} shows how the response function can be used to translate changes in the ionising continuum into corresponding emission line changes. In the limit of small variations in the ionising continuum luminosity, the response function $\Psi_{R}$ is constant. In this case, the time-dependent emission line, $L(v, t)$, can be expressed straightforwardly as

\begin{equation}
	L(v,t) = L_0(v)+\int^{\infty}_0 \Psi_R(v, \tau) \, \Delta C(t-\tau) \, d\tau,
	\label{eqn:method:basic}
\end{equation}
where $L_0(v)$ is the base-line reference spectrum (c.f. Equation~\ref{eq:linearize}). This is sufficient to generate spectra for CARAMEL which only requires continuum-subtracted line spectra. One key difference, however, is that this equation makes the assumption that the response is linearized, whereas CARAMEL itself does not. \memecho\ requires a full spectrum including large regions of continuum either side of the line. Adding continuum variations to Equation \ref{eqn:method:basic} results in

\begin{equation}
	L(v,t) = L_0(v)+C_0(v)+\Delta C(v, t)+\int^{\infty}_0 \Psi_R(v, \tau) \, \Delta C(t-\tau) \, d\tau.
	\label{eqn:method:memecho}
\end{equation} We generate our time series from these equations using the response functions discussed in Section~\ref{sec:method:fundamentals:response} and a variable driving continuum $C(t)$ based on the empirical 1158~\AA\ light-curve of NGC~5548 from \citet{DeRosa2015a}. The light-curve is rescaled to match the mean luminosity for each model, and the range of variation is reduced to $\pm50\%$ about this mean value. The data set contains 171 observations spread over $\simeq 175$~days. Given a BLR extending out to $R_{\rm max}$, the line profile at time $t$ will depend on continuum levels as far back as $t - 2R_{\rm max}/c$. It is therefore not possible to calculate self-consistent line profiles for times earlier than $2R_{\rm max}/c$ in the time series, since these depend on unknown continuum values. This affects not only our ability to simulate spectra, but any attempt to invert observational spectroscopic time-series. Thus, we discard $L(v,t)$ calculated at early times, when the line profiles still depend on unobserved continuum fluxes. We do, however, retain the continuum values observed during these early times. We thus provide both \memecho\ and CARAMEL with 171 continuum measurements across 175 days, but only 101 simultaneous line plus continuum spectra over 100 days. This is comparable to the best existing observation campaigns \citep{Du2014, DeRosa2015}. The spectra are given a constant error such that the error on the integrated line flux measured from a single spectrum is, on average, 2\% of the peak-to-peak variation in the integrated line flux, as requested by the CARAMEL team. These errors are also used to apply noise to our simulated time series. Representative MCMC samples of the line profiles from one of our simulated data sets are shown shown in Figure~\ref{fig:results:lightcurve_spectra}. The full time-series are shown in the form of trailed spectrograms in Figure~\ref{fig:method:tss}. The method used in the generation of time series is described in full in the the appendix to this work.

\subsection{Benchmarks: Defining Success}
\label{sec:method:benchmarks}

To assess the results produced by the RM techniques we are testing, we need to define what constitutes success. At the most basic level, any successful RM analysis must be consistent with the input data. Thus quantities like the mean line profiles, response functions and full spectroscopic time-series should be well reproduced by whatever inversion method is being used. Unless this minimal requirement is met, it is impossible to have confidence in the results obtained or their interpretation. The relevant benchmarks for our input models have already been shown and discussed in Sections~\ref{sec:method:fundamentals:line} (mean line profiles: Figure~\ref{fig:method:profile}), \ref{sec:method:fundamentals:response} (response functions: Figures~\ref{fig:method:response_sey} and \ref{fig:method:response_qso}), and \ref{sec:method:tss} (spectroscopic time series: Figure~\ref{fig:method:tss}).

\begin{figure*}
	\begin{minipage}{.48\textwidth}
      \includegraphics[width=\columnwidth]{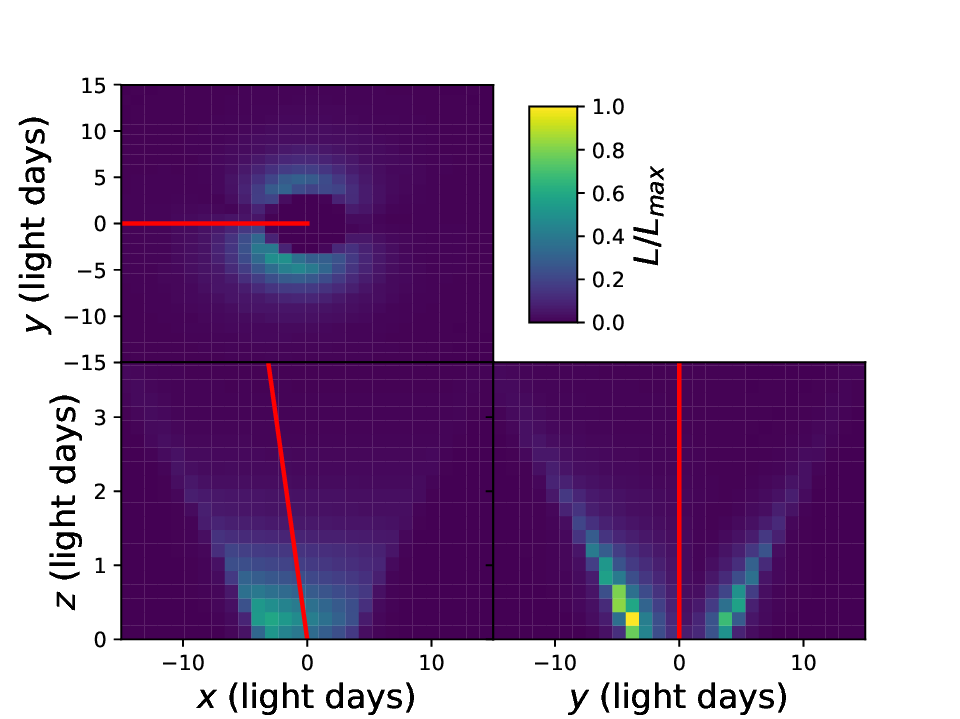}
      \caption{The emissivity distribution in the QSO model. Distances have been rescaled to correspond to the rescaled delays (see \ref{sec:method:fundamentals:response}). X and Y axes are along the disk plane, Z is normal to the disk plane. The red lines indicate the projection of the direction vector towards the observer in each plot. Note the different (smaller) dynamic range used for the z-axis.}
      \label{fig:geometry-emissivity}
	\end{minipage}\hfill\begin{minipage}{.48\textwidth}
      \includegraphics[width=\columnwidth]{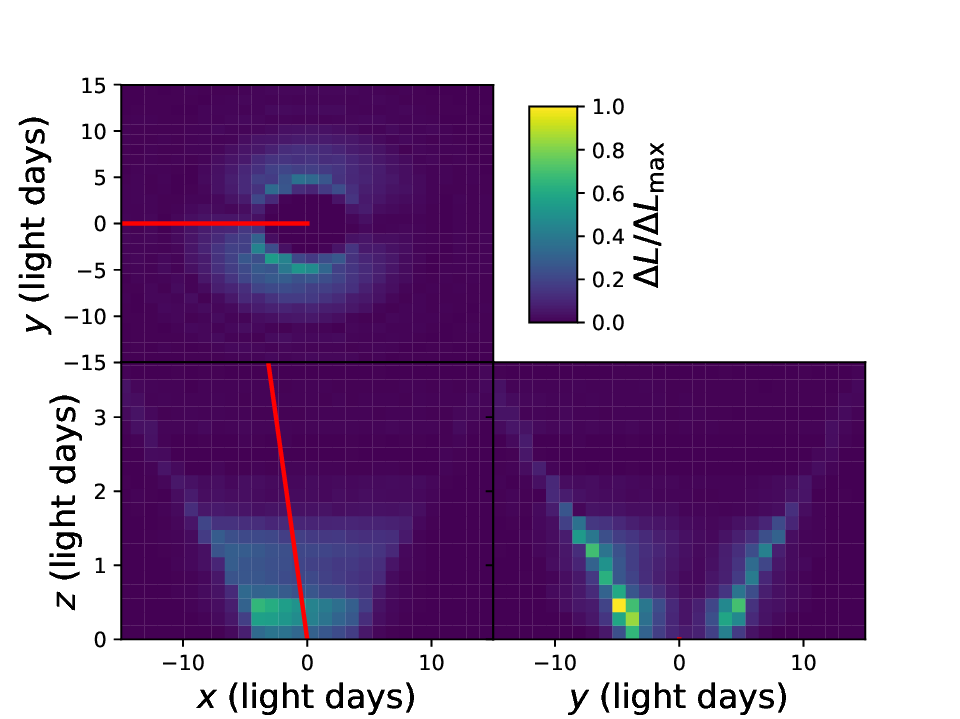}
      \caption{The responsivity-weighted emissivity distribution in the QSO model. X and Y axes are along the disk plane, Z is normal to the disk plane. Distances have been rescaled to correspond to the rescaled delays (see \ref{sec:method:fundamentals:response}). The red lines indicate the projection of the direction vector towards the observer in each plot. Note the different (smaller) dynamic range used for the z-axis.}
      \label{fig:geometry-responsivity}
    \end{minipage}
\end{figure*}

Of course, the real goal of RM is to gain insight into the physical nature of the BLR. Success in this context means correctly inferring physical properties such as the characteristic size of the line-forming region, its geometry, and the dominant kinematics. As additional benchmarks for assessing performance in this area, we provide in Figures~\ref{fig:geometry-emissivity} and ~\ref{fig:geometry-responsivity} the spatially resolved ``raw" and responsivity-weighted emissivity distributions for our QSO model.
\footnote{The corresponding emissivity distributions for the Seyfert model have been omitted here,  since neither of the RM methods we tested was able to reproduce the negative response of this model (see Section~\ref{sec:results:seyfert}).} Both of these matter. The former controls the shape of the {\em mean} line profile, while the latter controls the velocity-dependent line response and the RMS line profile.

The raw and responsivity-weighted emissivity distributions illustrate where in our biconical disc wind model the simulated H$\alpha$ line is primarily formed, and which parts of the line-emitting region are most sensitive to changes in the continuum. Note that there is no significant line emission from $z < 0$ (i.e. below the disk plane), since the optically thick accretion disc blocks the observer's view of this region. In addition, even though the line-forming region is vertically extended, its aspect ratio is small, $H/R \sim 0.1$ (note the different scales on the axes of Figures~\ref{fig:geometry-emissivity} and \ref{fig:geometry-responsivity}). Thus, geometrically, the H$\alpha$ line in this model could be reasonably described as being formed in a moderately thin disc or annulus, extending over 3 to 6 light days.

The emission distribution in Figure \ref{fig:geometry-emissivity} differs from that shown in \citet{Murray1996}. This is an orientation effect; when observed at a $40^\circ$ angle instead of $87^\circ$, the observer-projected component of the outflow is consistently positive, and acts to suppress the quadripolar $dv/ds$ distribution.

\subsection{Blinding}
\label{sec:method:blinding}

In order to ensure that our RM tests are realistic, we carried them out as blinded trials. Thus neither P.W. and A.P. (the CARAMEL team), nor K.H. (the one-man \memecho\ team) were given prior access to the response functions used to generate the time-series. They were also not given the disc wind model parameters we adopted. Instead, both were provided only with the time-series for the QSO and Seyfert models in their preferred input format, as well as the rescaled continuum light curves used to generate them. Neither were informed that the Seyfert model would exhibit a negative response.

The following methods sections for each technique were written by their respective teams, after they had access to the time-series, but before they were shown the actual \textsc{Python}-generated response functions that were used to produce the data.

One crucial difference to note between the methods is that, like our method, \memecho\ assumes a linearized response around a mean line profile as Equation~\ref{eq:linearize} and the velocity-delay maps it generates represent the \emph{response} function of the system. CARAMEL, however, does \emph{not} assume linearization, as Equation~\ref{eqn:background:1d_rf}, and the velocity-delay maps it generates are \emph{transfer} functions. This difference is important as the time-series of spectra were generated under the assumption of a \emph{linearized} response around a mean line profile.

\begin{figure}
	\includegraphics[width=\columnwidth]{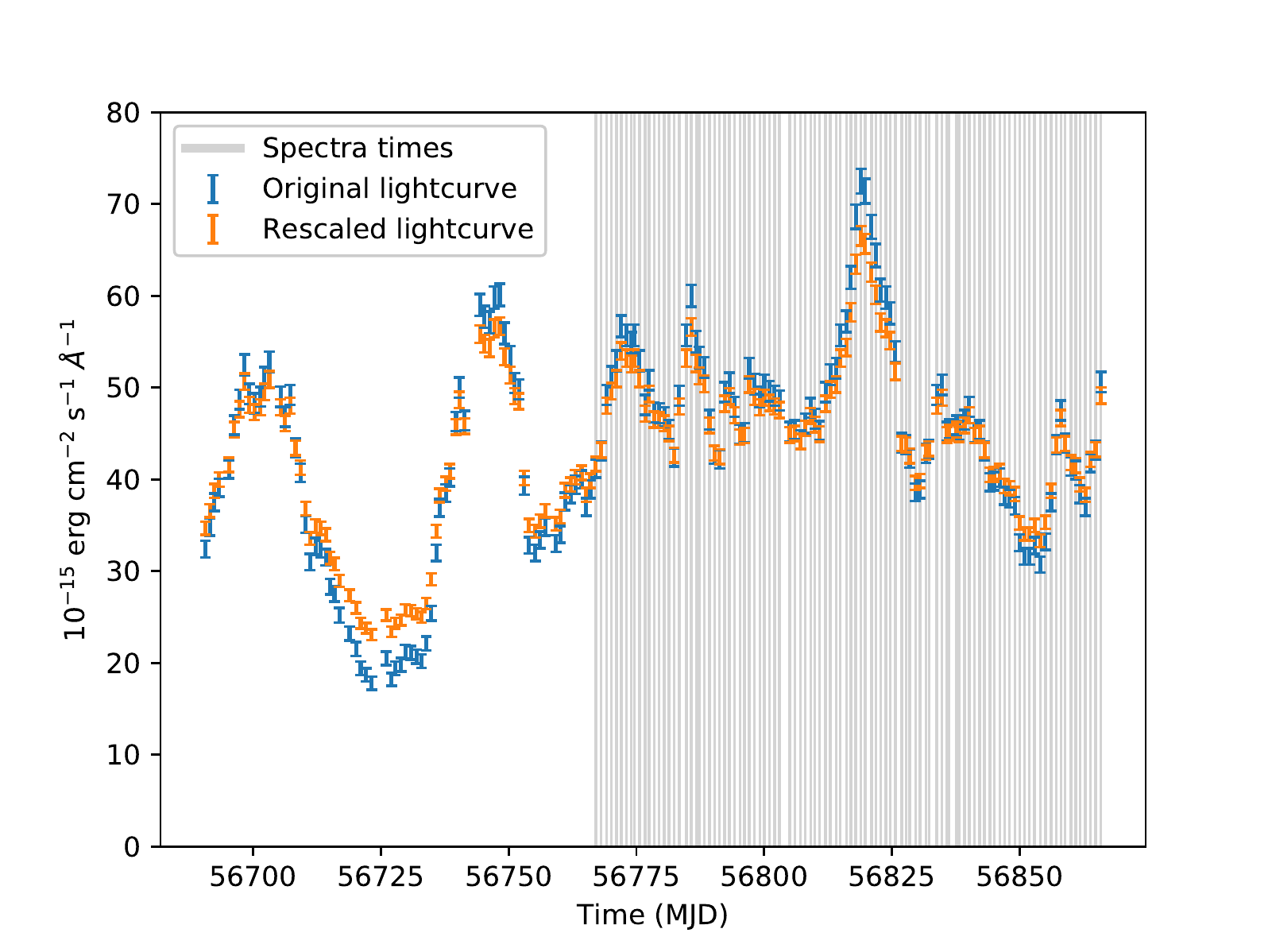}
	\caption{Original and rescaled driving light-curves used in generating time series of spectra, taken from NGC~5548 \citep{Fausnaugh2016}.}
    \label{fig:method:lightcurve}
\end{figure}

\subsection{Inversion Methods: \memecho\ }
\label{sec:method:memecho}
\label{sec:1d}
\typeout{1d}
We interpret the observed spectral variations
as time-delayed responses to a driving light-curve.
By fitting a model to the reverberating spectrum $F(\lambda,t)$,
we reconstruct a 2-dimensional wavelength-delay map $\Psi_R(\lambda,\tau)$.
This effectively ``slices up'' the accretion flow on isodelay surfaces,
which are paraboloids co-axial with the line of sight with a
focus at the compact source.
Each delay slice gives the spectrum of the response,
revealing the fluxes and Doppler profiles of emission lines
from gas located on the corresponding isodelay paraboloid.
The resulting velocity-delay maps $\Psi_R(v,\tau)$ provide
2-dimensional images of the accretion flow, one for each emission line,
resolved on isodelay and iso-velocity surfaces.

\subsubsection{ \memecho\ fits to the synthetic data }

In our blind analysis,
we treat the synthetic spectra in exactly the same way
as in the analysis of time-resolved spectroscopy of real AGN.
Thus a linear continuum model fit (using the linearized model as summarised in Equations~\ref{eqn:background:cont_lin}, \ref{eq:linearize} \& \ref{resp10}) to each of the synthetic
spectra provides the continuum light-curve data,
and the continuum-subtracted spectra isolate H$\alpha$
light-curve data in many wavelength channels. Note that \memecho\ did not use the full 171 continuum measurements including the preceding times; only those corresponding to the times for which spectra were available.

Using the \memecho\ code \citep{Horne1994} we then perform
regularised fits of the linearised echo model,
with parameters $p_k$, to the synthetic data $D_i\pm\sigma_i$.
The data $D$ comprise measurements at specific times
of the continuum light-curve and of the emission-line flux in each wavelength channel.
For 1-D echo mapping, the parameters $p$ include 3 parts:
the continuum light-curve $C(t)$, the echo map $\Psi_R(\tau)$,
and the line reference level $L_0$.
$C(t)$ and $\Psi_R(\tau)$ are evaluated on suitable
time and delay grids, with equal spacing $\Delta t=1$ day,
interpolating as needed to match the observation times.
We set the continuum reference level $C_0$ to the median of
the continuum data, and the fit adjusts $L_0$ accordingly.
For 2-D velocity-delay mapping, the data $D$ comprise
the continuum light-curve plus H$\alpha$ light-curves in many
wavelength channels, and the
model parameters include $\Psi_R(\lambda,\tau)$ and $L_0(\lambda)$
in the same wavelength channels.

Our \memecho\ fit is achieved by varying the model parameters $p$ to minimise
\begin{equation}
	Q(p,D) = \chi^2(p,D) - 2\, \alpha\, S(p)
\ .
\end{equation}
Here the ``badness-of-fit'' statistic
\begin{equation}
	\chi^2 = \sum_{i=1}^N \left( \frac{ D_i - \mu_i(p) } { \sigma_i } \right)^2
\end{equation}
quantifies consistency between the linearised echo model predictions $\mu_i(p)$
and the data values $D_i\pm\sigma_i$.
The fit employs a Bayesian prior $\propto \exp{\left\{\alpha\,S(p)\right\}}$,
where the entropy is
\begin{equation}
	S(p) = \sum_k w_k\, \left\{\, p_k - q_k - p_k\, \ln{\left( p_k / q_k \right)} \,\right\}
\ ,
\end{equation}
where $w_k$ is the weight and $q_k$ is the default value of parameter $p_k$.
Note that $S(p)$ requires $p_k>0$ and that
\begin{equation}
	\frac{\partial Q}{\partial p_k}
	= 2 \sum_{i=1}^N \frac{ D_i - \mu_i(p) }{ \sigma_i^2 }
	\frac{\partial \mu_i}{\partial p_k }
	+ 2\, \alpha\, w_k\, \ln{ \left( p_k / q_k \right) }
\ ,
\end{equation}
so that as $\chi^2$ pulls the model prediction $\mu_i(p)$ toward the data $D_i$,
$\alpha\, S$ pulls each parameter $p_k$ toward its default value $q_k$.
The default values are set to weighted averages of ``nearby'' parameters,
e.g. $q(t)=\sqrt{ p(t-\Delta t)\,p(t+\Delta t) }$,
so that the entropy favours smoothly-varying functions $C(t)$ and $\Psi_R(\tau)$.
We also ``pull down'' on $\Psi_R(\tau)$ at the maximum delay $\tau_{\rm max}=30$~d.
The weights $w_k$ depend on control parameters.
Increasing the control parameter $W$ ``stiffens'' structure
in $\Psi_R(\tau)$ relative to that in $C(t)$.
Similarly, in 2-D velocity-delay mapping, a second parameter $A$ controls the
trade-off in $\Psi_R(\lambda,\tau)$ between structure in the delay vs wavelength direction,
and parameter $B$ controls ``stiffness'' of $L_0(\lambda)$.

In practice we use \memecho\ to follow a maximum entropy trajectory from an initial large
value of $\alpha$ and corresponding large $\chi^2$, decreasing $\alpha$ and thus $\chi^2$
gradually until an ``acceptable'' $\chi^2$ is reached.
A measure of the angle between the gradients of $S$ and $\chi^2$ provides
evidence that the fit at each chosen $\chi^2$ level is converged to machine precision.
A value near $\chi^2/N=1$ is expected for $N$ data values with reliable
error bars when the linearised echo model can achieve an acceptable fit.
Attempts to lower $\chi^2$ too far result in a dramatic increase in $S$
as the parameters become noisy due to over-fitting noise in the data.
From the resulting series of \memecho\ fits we choose the
control parameters $W$ and $A$ and the $\chi^2$ level so as to achieve
plausible fits that reproduce the data well while keeping
relatively smooth the continuum light-curve $C(t)$ and response maps $\Psi_R(\lambda,\tau)$.

\subsection{Inversion Methods: CARAMEL}
\label{sec:method:caramel}

CARAMEL produces transfer functions from data by directly modelling the broad emission line region as a distribution of many massless point particles surrounding an ionizing continuum source.
The particles instantaneously and linearly reprocess the AGN continuum emission and re-emit the light towards the observer in the form of emission lines.
Each particle re-emits at a wavelength determined by its line-of-sight velocity and with a time-lag determined by its position.
Using the particles' positions and velocities, CARAMEL can then calculate the transfer function resulting from a given BLR model.

The model consists of a geometric component describing the positions of the point particles and a dynamical component describing the particle velocities.
In addition, CARAMEL models the continuum light curve using Gaussian processes as a flexible interpolator to produce simulated spectra at arbitrary times.
By feeding the continuum light curve through the BLR model, the code produces a time-series of spectra that can be directly compared to data.
We use a Gaussian likelihood function to compare the model to the data, and use the diffusive nested sampling code {\sc DNest3} \citep{Brewer2011} to explore the parameter space of the BLR and continuum models.
The full details of CARAMEL and the BLR model are discussed by \citet{Pancoast2014}, but the main components are described in the rest of this section.

The particles in the BLR model are first assigned radial positions drawn from a Gamma distribution,
\begin{align}
	p(x | \alpha, \theta) \propto x^{\alpha - 1} \exp\left(-\frac{x}{\theta}\right)
\end{align}
which allows for Gaussian-like, exponential, and heavy-tailed distributions.
The distribution is then offset from the origin by the Schwarzschild radius plus a minimum BLR radius, $r_{\rm min}$, and a change of variables between ($\alpha$, $\theta$, $r_{\rm min}$) and ($\mu$, $\beta$, $F$) is applied:
\begin{align}
	\mu &= r_{\rm min} + \alpha \, \theta \\
	\beta &= \frac{1}{\sqrt{\alpha}} \\
	F &= \frac{r_{\rm min}}{r_{\rm min} + \alpha \, \theta},
\end{align}
where $\mu$ is the mean radius, $\beta$ is the shape parameter, and $F$ is the minimum radius in units of $\mu$.
In this formalism, the standard deviation of the Gamma distribution is $\sigma_r = \mu\beta(1 - F)$.
The particles are then rotated out of the plane of the disc by a random angle uniformly distributed over the range $\pm\theta_o$, and the distribution is inclined by an angle $\theta_i$ relative to the observer, where $\theta_{i} \rightarrow 0^\circ$ is face-on.
The opening angle prior $\theta_0$ is uniform between $0^\circ$ and $90^\circ$, and the inclination angle prior is uniform in $\cos\theta_i$ between $0^\circ$ and $90^\circ$

The emission from each particle is assigned a weight between $0$ and $1$, determined by
\begin{align}
	W(\phi) = \frac{1}{2} + \kappa \, \cos\phi,
\end{align}
\label{eq:caramel_weight}
where $\phi$ is the angle from the observer's line of sight to the origin to the particle position, and $\kappa$ is a free parameter with uniform prior between $-0.5$ and $0.5$.
In this set-up, $\kappa = 0$, $-0.5$, and $0.5$ correspond to particles that emit isotropically, back towards the ionizing source, and away from the ionizing source, respectively.
A parameter $\gamma$, with uniform prior between 1 and 5, allows the particles to be distributed uniformly throughout the BLR ($\gamma\rightarrow 1$) or clustered near the faces of the disc ($\gamma\rightarrow 5$).
This is achieved by setting the angle between a point particle and the disc to be
\begin{align}
	\theta = \arccos \left[\cos \theta_o + (1 - \cos\theta_o) \, U^\gamma\right],
\end{align}
where $U$ is drawn randomly from a uniform distribution between 0 and 1.
Finally, an additional free parameter, $\xi$ (uniform prior between $0$ and $1$), allows the disc mid-plane to be opaque ($\xi\rightarrow 0$) or transparent ($\xi\rightarrow 1$).

The wavelength of light emitted by each particle is determined by its velocity, which is in turn determined by the black hole mass, a free parameter with uniform prior in the log of $M_{\rm BH}$ between $2.78\times 10^4$ and $1.67\times 10^9$ $M_\odot$, and the parameters $f_{\rm ellip}$, $f_{\rm flow}$, and $\theta_e$.
First, the particles are assigned to be on near-circular elliptical orbits or on either inflowing or outflowing orbits.
The fraction of particles on near-circular orbits is determined by the free parameter $f_{\rm ellip}$ which has a uniform prior between 0 and 1.
Those with near-circular elliptical orbits have their radial and tangential velocities drawn from a Gaussian distribution centred on the circular velocity in the $v_\phi - v_r$ plane.
The remaining particles are drawn from Gaussian distributions centred on the radial inflowing or outflowing escape velocities, where the direction of flow is determined by the parameter $f_{\rm flow}$.
$f_{\rm flow}$ has a uniform prior between 0 and 1, where $f_{\rm flow} < 0.5$ ($>0.5$) indicates inflow (outflow).
We also allow the centres of the inflowing and outflowing distributions to be rotated by an angle $\theta_e$ towards the circular velocity in the $v_\phi - v_r$ plane, where $\theta_e$ has a uniform prior between $0^\circ$ and $90^\circ$.

\section{Results and Discussion}
\label{sec:results}

In the following sections, we will present the blinded analysis and interpretation of the simulated QSO data, as obtained by the two methods and written by their respective teams (\memecho\: Section~\ref{sec:results:memecho}; CARAMEL: \ref{sec:results:caramel}). We will then unblind the analysis and compare the blind results to ``ground-truth", i.e. to the known response function and the underlying QSO BLR model (\ref{sec:results:models}).

First, however, in Section~\ref{sec:results:seyfert}, we briefly consider the results obtained for the Seyfert data set. Perhaps unsurprisingly, both methods struggled to deal with the negative line response exhibited by this model. Since neither method obtained acceptable fits to this data set, it does not make sense to force a detailed analysis and interpretation of the ``best" models. If similarly poor fits were obtained for actual data, we would expect this to be interpreted (correctly) as evidence of a mis-specified model. We will therefore simply summarize the difficulties encountered by both deconvolution methods when faced with this model.

\subsection{\memecho\ and CARAMEL Results for the Seyfert model}
\label{sec:results:seyfert}

Neither \memecho\ nor CARAMEL were able to successfully fit our simulated Seyfert data set. In both cases, the underlying problem is the negative line response presented by this model. On the one hand, it is reassuring that neither method was ``fooled" by this data set -- i.e. both methods failed, rather than producing misleading results. On the other hand, the inability of both methods to deal with negative emission line responses is a significant limitation. In particular, it is unclear whether this could cause serious systematics in cases where the {\em net} response is clearly positive, but where specific {\em parts} of the BLR exhibit negative responsivity. Answering this question is beyond the scope of the present paper, but must be the focus of future work. It is also worth noting that there {\em are} deconvolution methods that explicitly allow for negative responsivities, such as regularized linear inversion \citep{Krolik1995a,Skielboe2015} and even an extension to \memecho\ discussed by \citet{Horne1994}.

\memecho\ was able to produce a response function for the Seyfert model (Figure \ref{fig:twodseymap}). As the \memecho\ code is not designed to model negative responses the response function can only display the region of negative response as simply zero response, but surprisingly the regions of \emph{positive} response are captured reasonably well. The comparatively fine features corresponding to the positive response from the far side of the inner disk are even reflected, albeit smoothed out by regularization. The response is also shifted to a lower delay- with the peak moved from $\approx 2$ days to $\approx 0$. Despite this, the response function recovered still matches the Keplerian envelope of a $10^{8} M_\cdot$ central mass well, very close to the $1.33^10{8} M_\cdot$ rescaled mass for this model.

\begin{figure}
	\includegraphics[width=0.9\columnwidth]{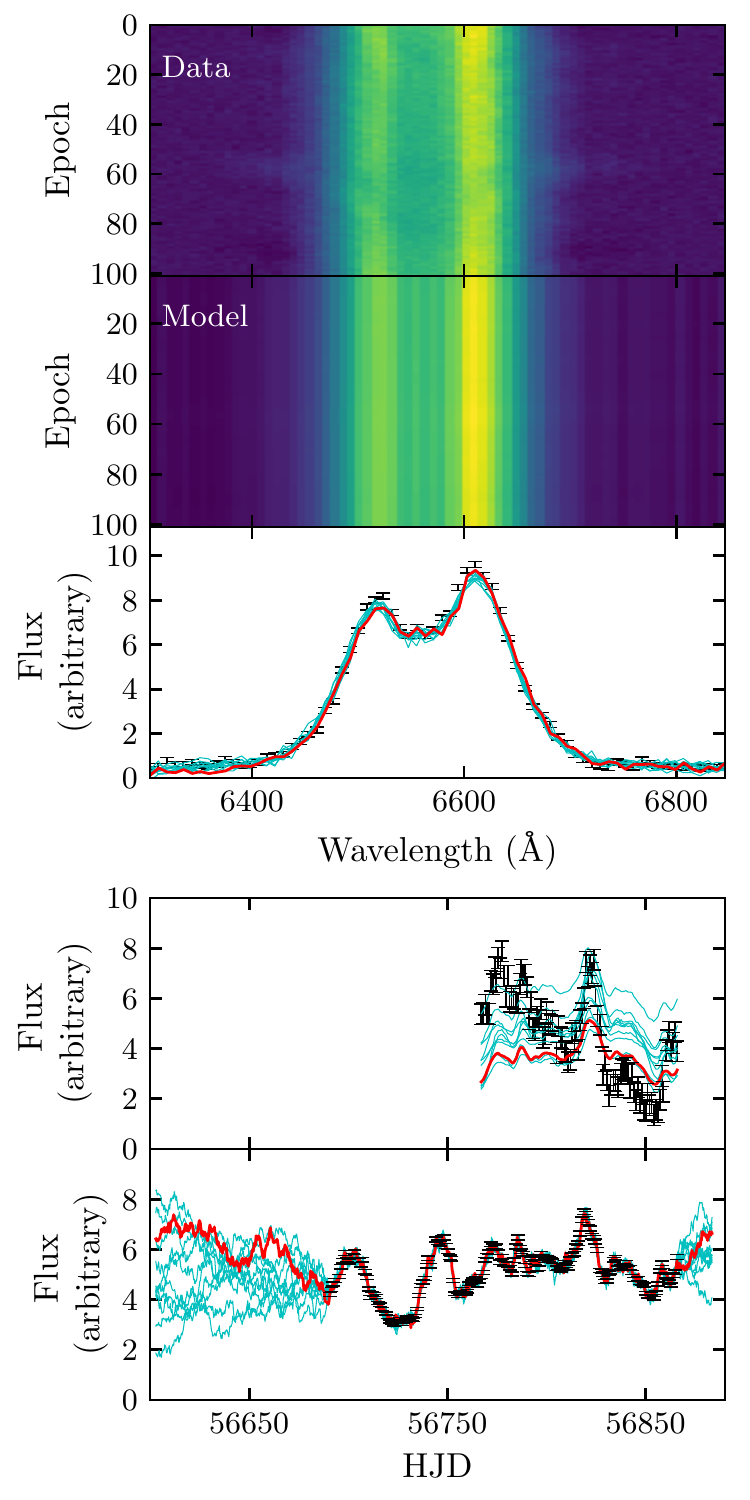}
    \caption{Model fits to the Seyfert model H$\alpha$ line profile, integrated H$\alpha$ flux, and AGN continuum flux.
	Panel $1$: The provided H$\alpha$ emission-line profile for each epoch.
	Panel $2$: The H$\alpha$ emission-line profile for each epoch produced by one sample of the BLR and continuum model.
	Panel $3$: The provided H$\alpha$ line profile for one randomly chosen epoch (black), and the corresponding profile (red) produced by the model in Panel 2. Cyan lines show the H$\alpha$ profile produced by other randomly chosen models.
	Panels $4$ and $5$: Time series of the provided integrated H$\alpha$ and continuum flux (black), the time series produced by the model in Panel 2 (red), and time series produced by other sample BLR and continuum models (cyan).
	}
    \label{fig:display_caramelsey}
\end{figure}

\begin{figure*}
\begin{center}
\includegraphics[angle=0,width=90mm]{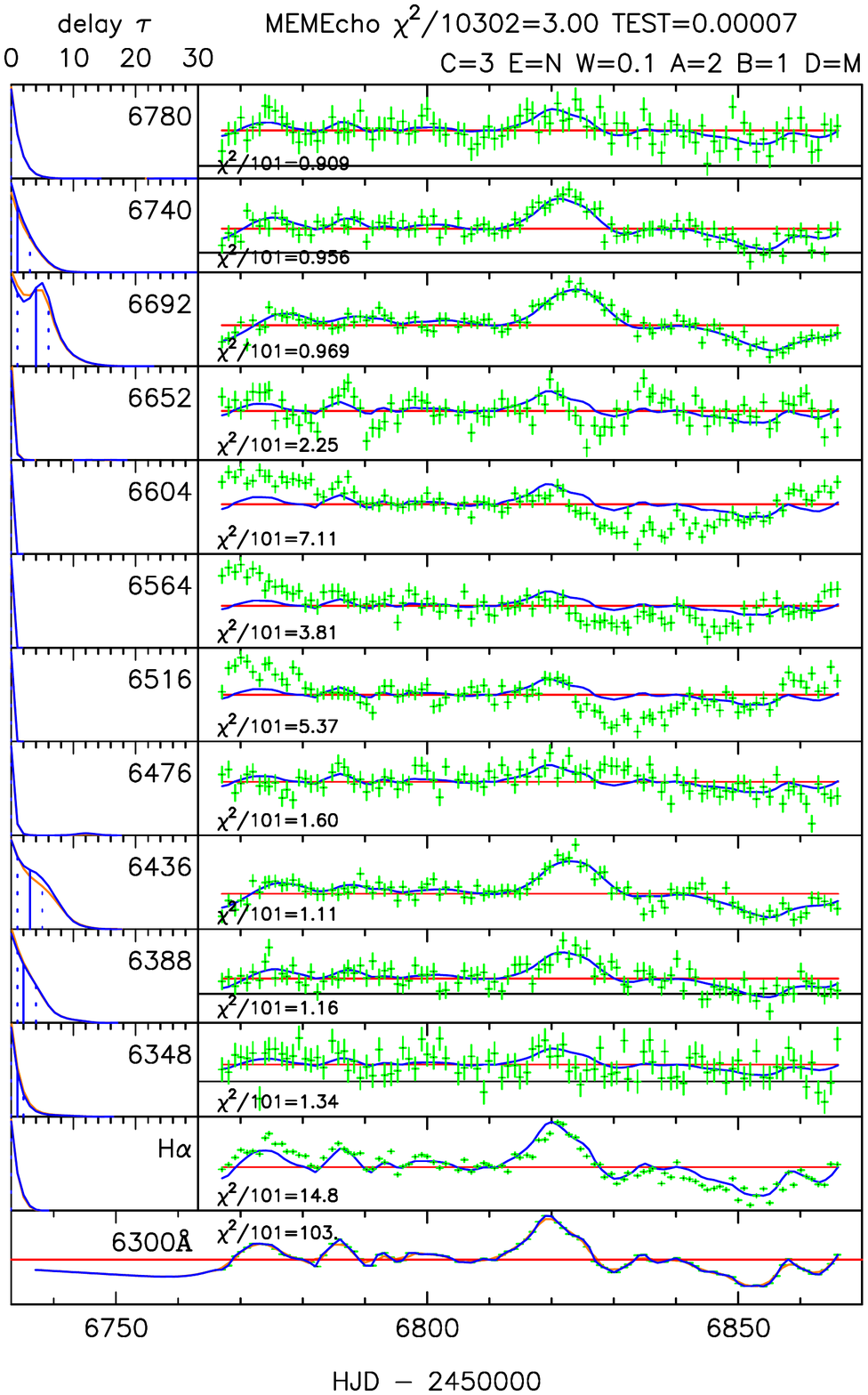}
\caption{
\label{fig:display_memechosey}
\memecho\ fit to the synthetic Seyfert data.
Bottom panel is the driving light-curve, above which
are the 1-D echo maps (left) and echo light-curves (right)
at selected wavelengths. Note the high target $\chi^2/101 = 3$ and the poor fits achieved near line center.
}
\end{center}
\end{figure*}

\begin{figure*}
\begin{center}
\includegraphics[angle=270,width=170mm]{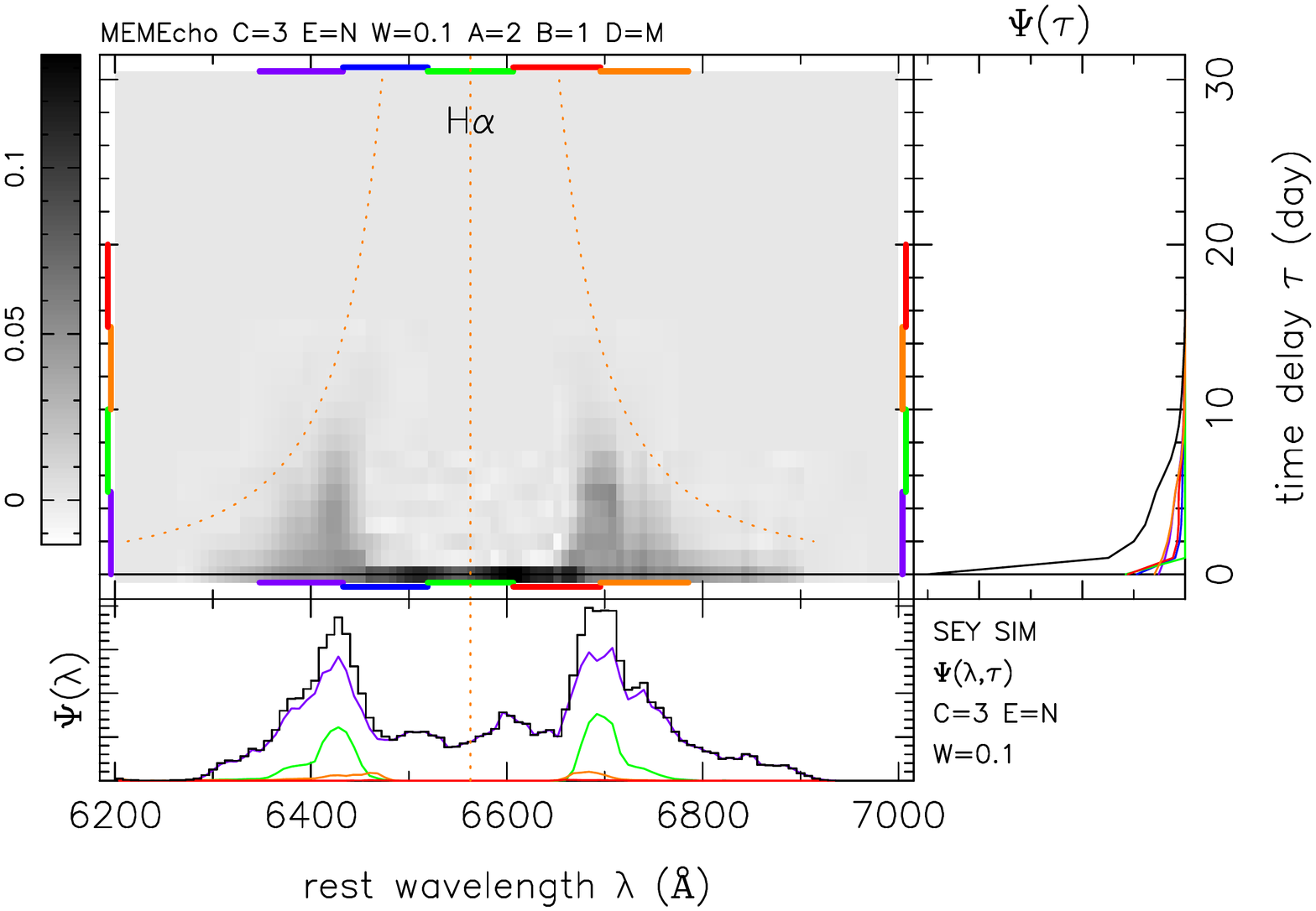}
\caption{ \label{fig:twodseymap}
Two-dimensional wavelength-delay map $\Psi_R(\lambda,\tau)$
reconstructed from the \memecho\ fit to the synthetic Seyfert data.
Given below the greyscale map are projections of $\Psi_R(\lambda,\tau)$
giving delay-integrated responses $\Psi_R(\lambda)$ for the full
delay range (black), and for restricted delay slices
0-5~d (purple), 5-10~d (green), 10-15~d (orange), and 15-20~d (red).
To the right of the greyscale map are wavelength-integrated
responses $\Psi_R(\tau)$
for the full line (black), and for 4000~km/s wide velocity
bins centred at $V=0$ (green), at $\pm4000$~\kms\  (red and blue),
and at $\pm8000$~\kms\ (orange and purple).
For a black hole mass of $\mbh=10^8\,\msun$,
the orange dotted curves show the virial envelope
for edge-on circular Keplerian orbits.
}
\end{center}
\end{figure*}

In order to provide some insight into how and why our benchmark methods struggle with the negative-response Seyfert model, we show in Figures~\ref{fig:display_caramelsey} and~\ref{fig:display_memechosey} a summary of the fits to this data set achieved by CARAMEL end \memecho, respectively. Taking the CARAMEL results first (Figure~\ref{fig:display_caramelsey}), we see that the overall {\em shape} of the H$\alpha$ line profile is reproduced very well, but that none of the models drawn from the posterior parameter distribution succeed in reproducing the integrated emission line light curve. In their interpretation of these results, the CARAMEL team correctly highlighted that this failure may be due to a non-linear response. Turning to \memecho\
(Figures~\ref{fig:display_memechosey} and \ref{fig:twodseymap}),
we first note that a high target $\chi^2/101 = 3$
had to be adopted in order to fit this data set. The resulting model reproduces well the light curve features in the wings of the emission line, but fails to reproduce the variations in the core of the H$\alpha$ line. Here, the model light curve is too low at the start and too high after the main peak at $t\approx6820$. This difficulty in matching the line center behaviour makes sense, since this is where the true response is most strongly negative. Both the CARAMEL and \memecho\ teams correctly interpreted the difficulties their methods encountered in fitting the Seyfert data set as pointing towards the presence of a significant negative response in the Seyfert model.

\subsection{Blind Analysis and Interpretation: \memecho\ Results for the QSO model}
\label{sec:results:memecho}

\subsubsection{1-D delay maps $\Psi_R(\tau)$ for the QSO simulation}

\begin{figure}
\begin{center}
\includegraphics[angle=0,width=90mm]{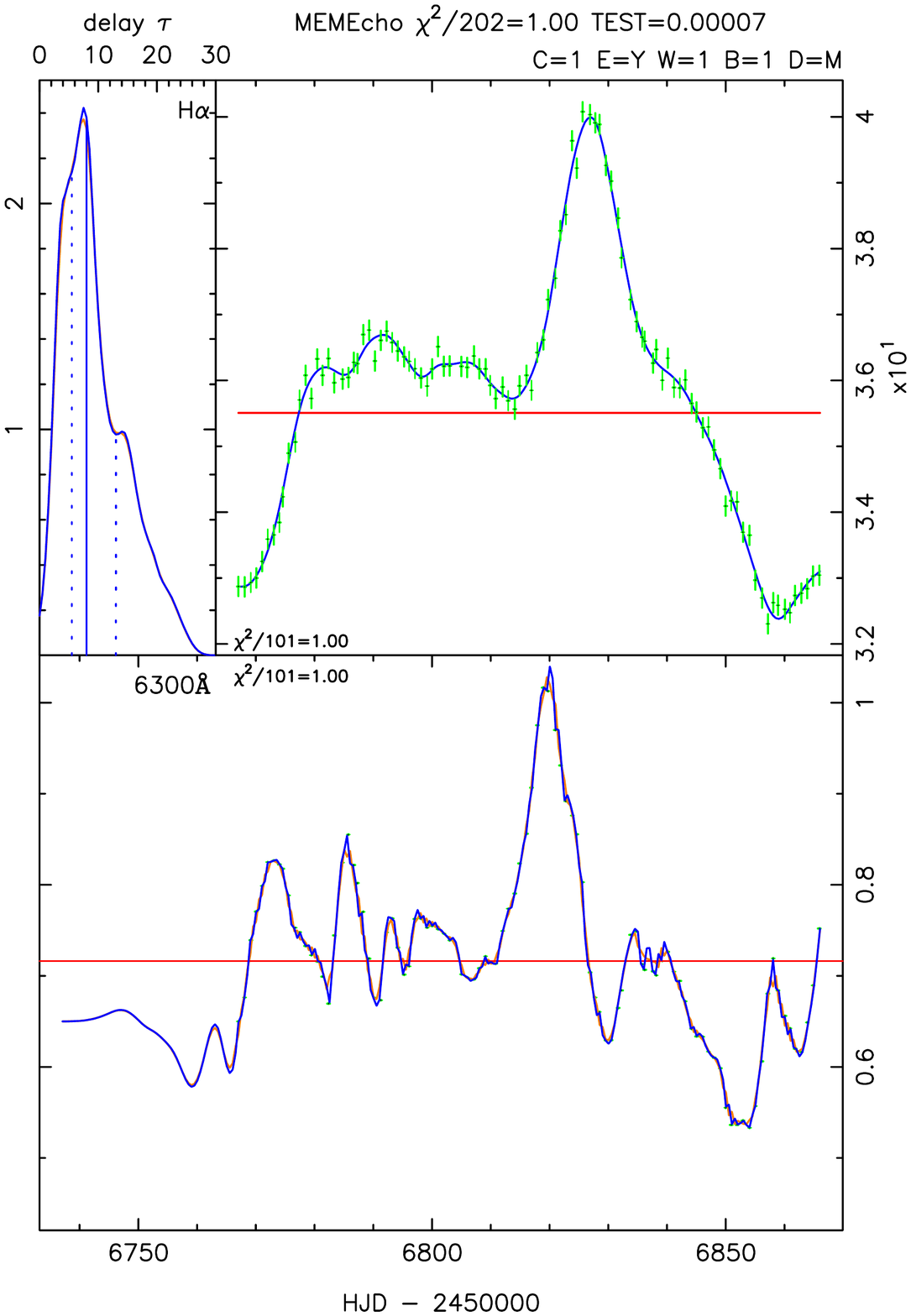}
\caption{ \label{fig:onedqsofit}
\memecho\ fit to the continuum and integrated H$\alpha$ light-curve
from the synthetic QSO dataset. The fit achieves $\chi^2/N=1$ both for the continuum variations
(lower panel) and the line variations (upper right panel). Blue curves show the fitted model, including
the continuum light-curve $C(t)$ (bottom panel), the delay map $\Psi_R(\tau)$ (upper left panel) and the
line light-curve $L(t)$ (upper right panel). Horizontal red lines indicate the reference levels $C_0$ for the continuum
and $L_0$ for the line.
}
\end{center}
\end{figure}

In Figure~\ref{fig:onedqsofit} we show the results of a successful \memecho\ fit to the synthetic QSO data.
The lower panel shows the continuum light-curve data, with error bars too small to see,
and in blue the fitted model driving light-curve $C(t)$.
The upper right panel shows the integrated line profile data
points with error bars (green) and the fitted line light-curve $L(t)$ (blue).
The reference levels, $C_0$ for the continuum and $L_0$ for the line,
are indicated by horizontal red lines.
The line variations $L(t)-L_0$ are obtained by convolving
the continuum variations $C(t)-C_0$ with the delay distribution $\Psi_R(\tau)$ shown in the upper left panel.
This fit was successful, achieving $\chi^2/N=1$ for the $N=101$
data points in the continuum light-curve and the same for the line light-curve.

The continuum light-curve has a large peak near $t=6820$ MJD,
and shows numerous smaller peaks and troughs that are well
detected at a high signal-to-noise ratio.
The strongest peak in the line light-curve crests at $t=6828$ MJD,
$\sim8$ days later than the corresponding peak in the continuum light-curve.
The line light-curve data is low near the start, then
rises to a plateau that has 3 or perhaps 4 local maxima before the main peak,
and drops to a low level again near the end.
Given the quality of the \memecho\ fit, these features can evidently be interpreted
in terms of the linear echo model.

 The 1-D delay map $\Psi_R(\tau)$ is well determined from the synthetic QSO data.
The prompt response at $\tau=0$ is small. The response rises rapidly
to a ledge near 4 days, a peak near 8 days, then declines to
a ledge at 17 days, and declines to near 0 at 30 days.
The data quality is high enough to warrant interpretation of these features.

\subsubsection{ 2-D Velocity-Delay Maps $\Psi_R(v,\tau)$ }
\label{sec:2d}
\typeout{2d}

\begin{figure}
\begin{center}
 \includegraphics[angle=0,width=90mm]{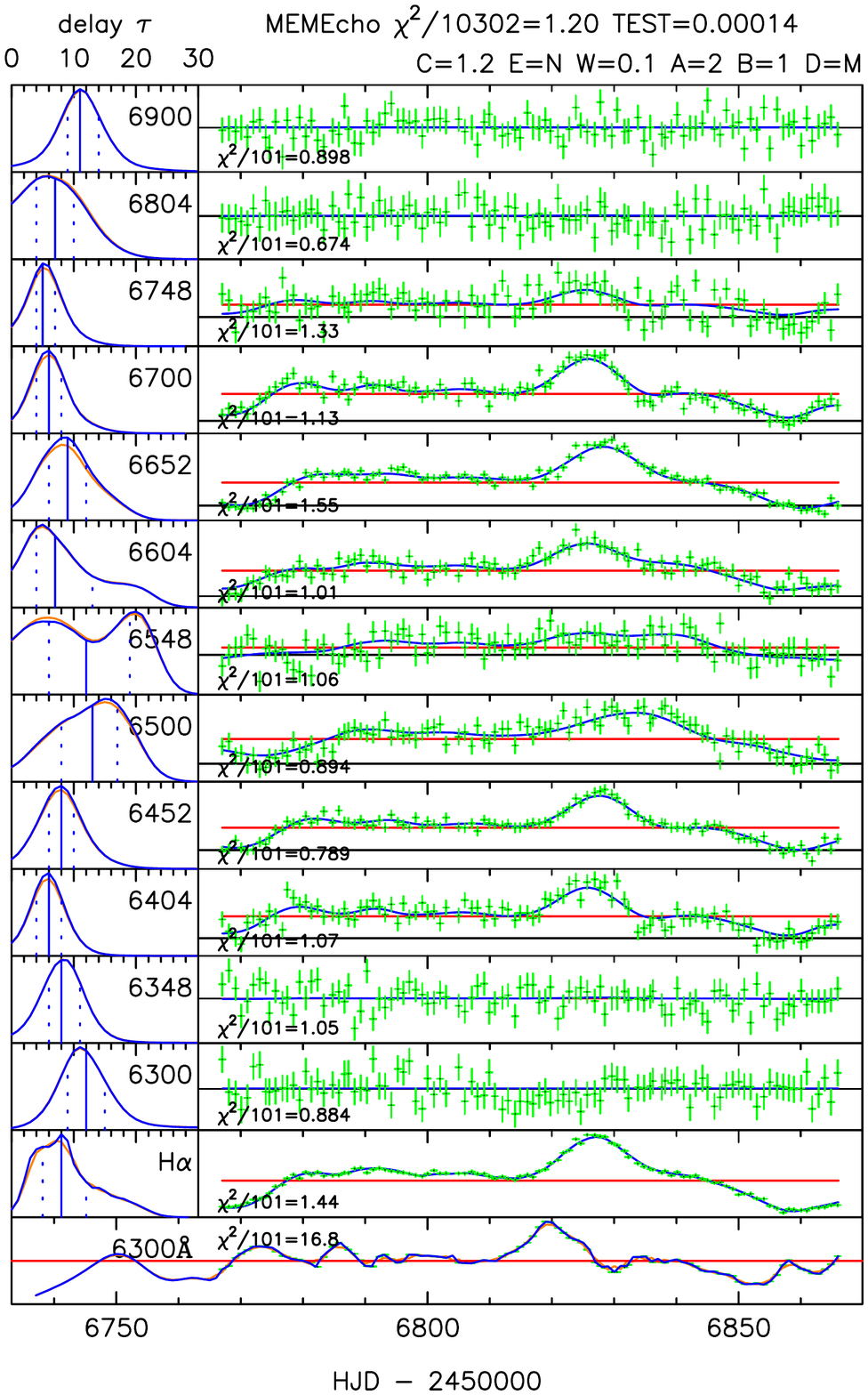}
\caption{ \label{fig:twodqsofit}
\memecho\ fit to the synthetic QSO data.
Bottom panel is the driving light-curve, above which
are the 1-D echo maps (left) and echo light-curves (right)
at selected wavelengths.
}
\end{center}
\end{figure}

\begin{figure*}
\begin{center}
 \includegraphics[angle=270,width=170mm]{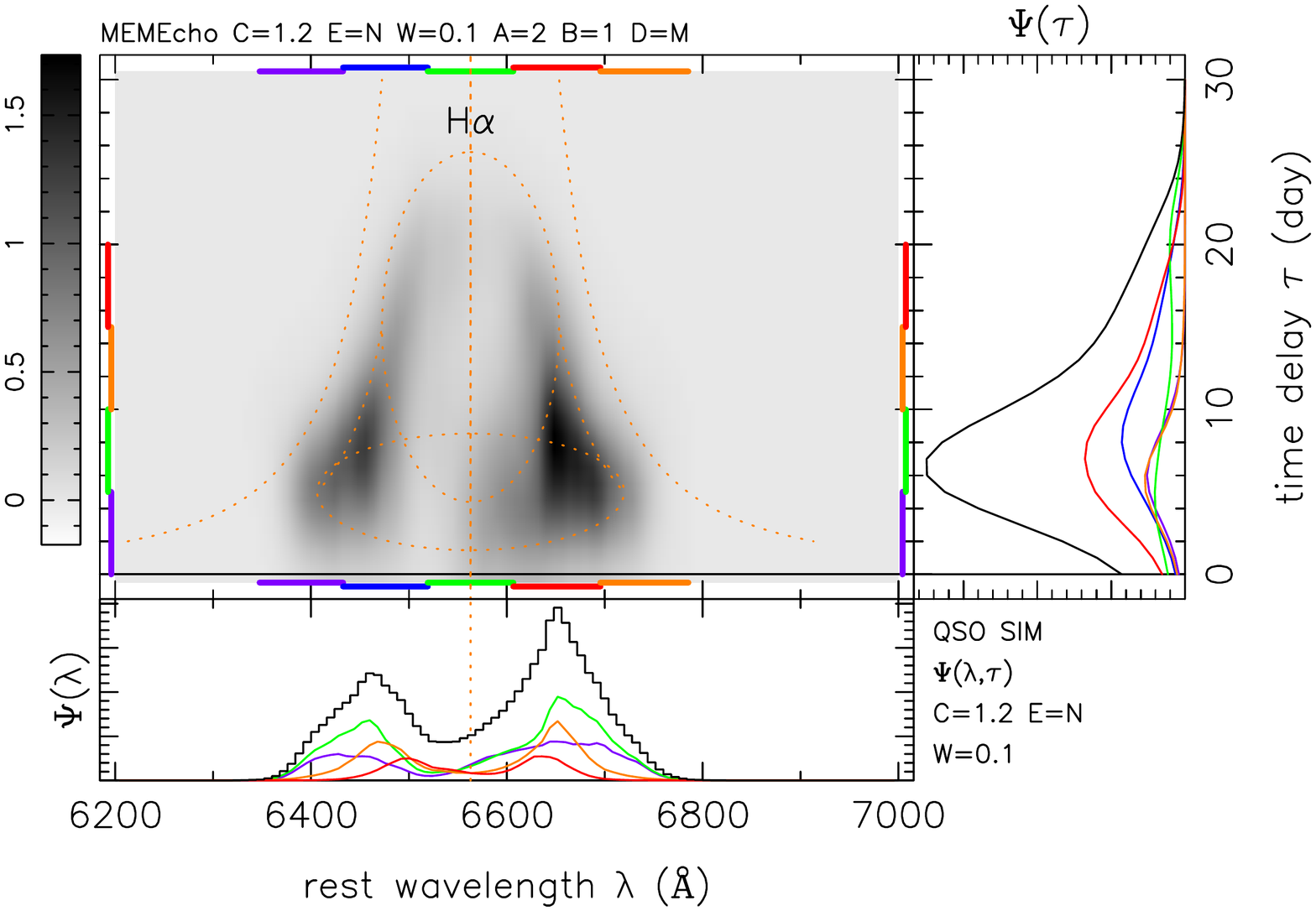}
\caption{ \label{fig:twodqsomap}
Two-dimensional wavelength-delay map $\Psi_R(\lambda,\tau)$
reconstructed from the \memecho\ fit to the synthetic QSO data.
Given below the grey-scale map are projections of $\Psi_R(\lambda,\tau)$
giving delay-integrated responses $\Psi_R(\lambda)$ for the full
delay range (black), and for restricted delay slices
0-5~d (purple), 5-10~d (green), 10-15~d (orange), and 15-20~d (red).
To the right of the grey-scale map are wavelength-integrated
responses $\Psi_R(\tau)$
for the full line (black), and for 4000~km/s wide velocity
bins centred at $V=0$ (green), at $\pm4000$~\kms\  (red and blue),
and at $\pm8000$~\kms\ (orange and purple).
For a black hole mass of $\mbh=10^8\,\msun$,
the orange dotted curves show the virial envelope
for edge-on circular Keplerian orbits, and for
a Keplerian disc inclined by $i=45^\circ$
and extending from $5$ to $15$ light days.
}
\end{center}
\end{figure*}

We now present the more detailed results of a 2-D \memecho\ fit to the synthetic QSO data
including variations not just in the continuum and integrated line flux,
but also in the emission-line velocity profile.

The quality of this \memecho\ fit may be judged from Figure~\ref{fig:twodqsofit}.
This shows the continuum light-curve and the integrated H$\alpha$ line light-curve
in the lower two panels.  Above these are delay maps on the left and echo light-curves
on the right for selected wavelengths as indicated.
The echo light-curve data, with green error bars, is shown along with the fitted
model $L(\lambda,t)$ (blue curve) and background level $L_0(\lambda)$ (red line).

The 2-D velocity-delay map $\Psi_R(\lambda,\tau)$ resulting from the
\memecho\ fit is shown as a grey-scale image in Figure~\ref{fig:twodqsomap}.
To the right are projections of $\Psi_R(\lambda,\tau)$
to form the velocity-integrated delay map $\Psi_R(\tau)$ (black)
and $\Psi_R(\tau|\lambda)$ for selected 4000~\kms\ wide velocity bins
centred at $V=0$ (green), $\pm4000$~\kms\ (orange and blue),
and $\pm8000$~\kms\ (red and purple).
Below the grey-scale map are projections of $\Psi_R(\lambda,\tau)$
to form the velocity profiles $\Psi_R(V)$
for the delay-integrated response (black),
and $\Psi_R(V|\tau)$, in colour, for 4 delay bins,
0-5~d (purple), 5-10~d (green), 10-15~d (orange) and 15-20~d (red).

This \memecho\ fit achieved an overall $\chi^2/N=1.2$ for $N=10302$ data values.
A lower $\chi^2$ could also be achieved but not without introducing small-scale
structure in the map, suggestive of over-fitting to noise in the data.
Maps with higher $\chi^2$ were also constructed, and give poorer fits
to the data while smearing out the structure seen in the map for $\chi^2/N=1.2$.
The map shown is thus a good compromise between noise and resolution.

The \memecho\ map exhibits interesting velocity-delay structure.
To first order, the H$\alpha$ response has a double-peaked velocity
structure with a delay structure that is symmetric on the red and
blue sides of the line profile.
In more detail, the response is stronger on the red side
than on the blue side.

In the lower panel of Figure~\ref{fig:twodqsomap},
the delay-integrated response $\Psi_R(V)$ (black)
extends to $\pm10000$~\kms\, with two
roughly triangular peaks cresting at $V\approx\pm4000$~\kms.
The red peak is stronger and sharper than the blue one, and the central minimum is at $-1500$~\kms.
The velocity profiles in different delay bins are also double-peaked.
The sharpness, velocity separation and red/blue asymmetry in strength of the peaks,
and the velocity at the minimum between the peaks, all change with the time delay.
In the 0-5 day bin (purple), the response extends to $\pm10000$~\kms\, with two smooth dome-shaped
peaks, stronger and broader on the red than the blue side, and with a minimum near $-2000$~\kms.
In the $5-10$~d bin (green), the response still covers $\pm10000$~\kms\, but the triangular
peaks have now appeared near $\pm4000$~\kms, and the central minimum moves redward to perhaps $-1600$~\kms.
In the $10-15$~d bin (orange), the response declines in the far wings, the sharp peaks remain
but move inward somewhat, and the central minimum moves redward to $+400$~\kms.
In the $15-20$~d bin (red), the wings decline further, the two peaks move together to $\pm3000$~\kms\,
and the central minimum is near $+1200$~\kms.

The two velocity peaks are separated by $\pm4000$~\kms\ at $\tau\sim15$~d delays, moving closer together at longer delays,
perhaps toward a merger at a maximum delay of $25$~d.
This structure suggests the top half of an elliptical ring feature, such as might arise from an annulus of orbiting gas
with $R\approx15$ light days and $V\,\sin{i}\approx4000$~\kms.
For an inclined circular Keplerian orbit, with $\tau = (R/c)\, \left( 1 + \sin{i} \cos{\theta} \right)$
and $V = V_{\rm Kep}\, \sin{i}\,\sin{\theta}$, the implied black hole mass is $M_{\rm BH} \sin^2{i} \sim 5\times10^7\,\msun$.
For an ellipse with maximum delay 25~d and mean delay 15~d, $1+\cos{i}\approx1.67$ and thus $i\approx45^\circ$.
For $i=45^\circ$, then $M_{\rm BH}\approx10^8\,\msun$.

For a $10^8\,\msun$ black hole, the orange dotted curves on Figure~\ref{fig:twodqsomap}
show the virial envelope for edge-on circular Keplerian orbits,
and for $45^\circ$ inclined orbits extending from 5 to 15 light days.
This framework captures much of the structure evident
in the velocity-delay map, and provides a reference against
which to consider evidence for departures from that simple model.
Note that the outer disc ring becomes indistinct at maximum delay
of $\sim25$~d, and its lower edge is also missing or obscured.
There appears to be low or negative response
at $-3000<V<-1000$~\kms\ and $\tau<10$~d.
Such response gaps may arise from azimuthal structure
on the ring, suppressing the line response at the
corresponding azimuth.

\subsection{Blind Analysis and Interpretation: CARAMEL Results for the QSO model}
\label{sec:results:caramel}

\begin{figure}
	\includegraphics[width=0.9\columnwidth]{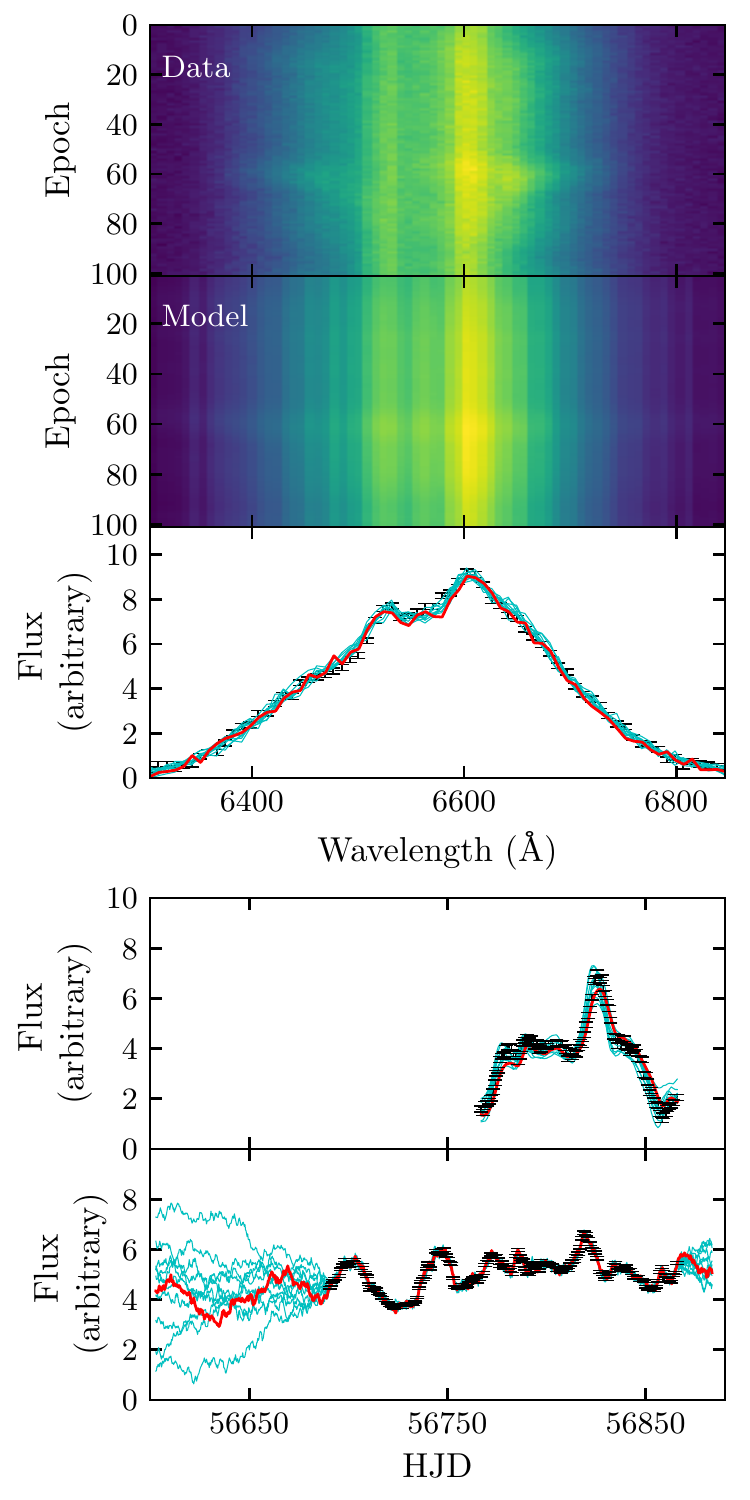}
    \caption{Model fits to the QSO model H$\alpha$ line profile, integrated H$\alpha$ flux, and AGN continuum flux.
	Panel $1$: The provided H$\alpha$ emission-line profile for each epoch.
	Panel $2$: The H$\alpha$ emission-line profile for each epoch produced by one sample of the BLR and continuum model.
	Panel $3$: The provided H$\alpha$ line profile for one randomly chosen epoch (black), and the corresponding profile (red) produced by the model in Panel 2. Cyan lines show the H$\alpha$ profile produced by other randomly chosen models.
	Panels $4$ and $5$: Time series of the provided integrated H$\alpha$ and continuum flux (black), the time series produced by the model in Panel 2 (red), and time series produced by other sample BLR and continuum models (cyan).
	}
    \label{fig:display_caramelqso}
\end{figure}

\begin{figure*}
	\includegraphics[width=\textwidth]{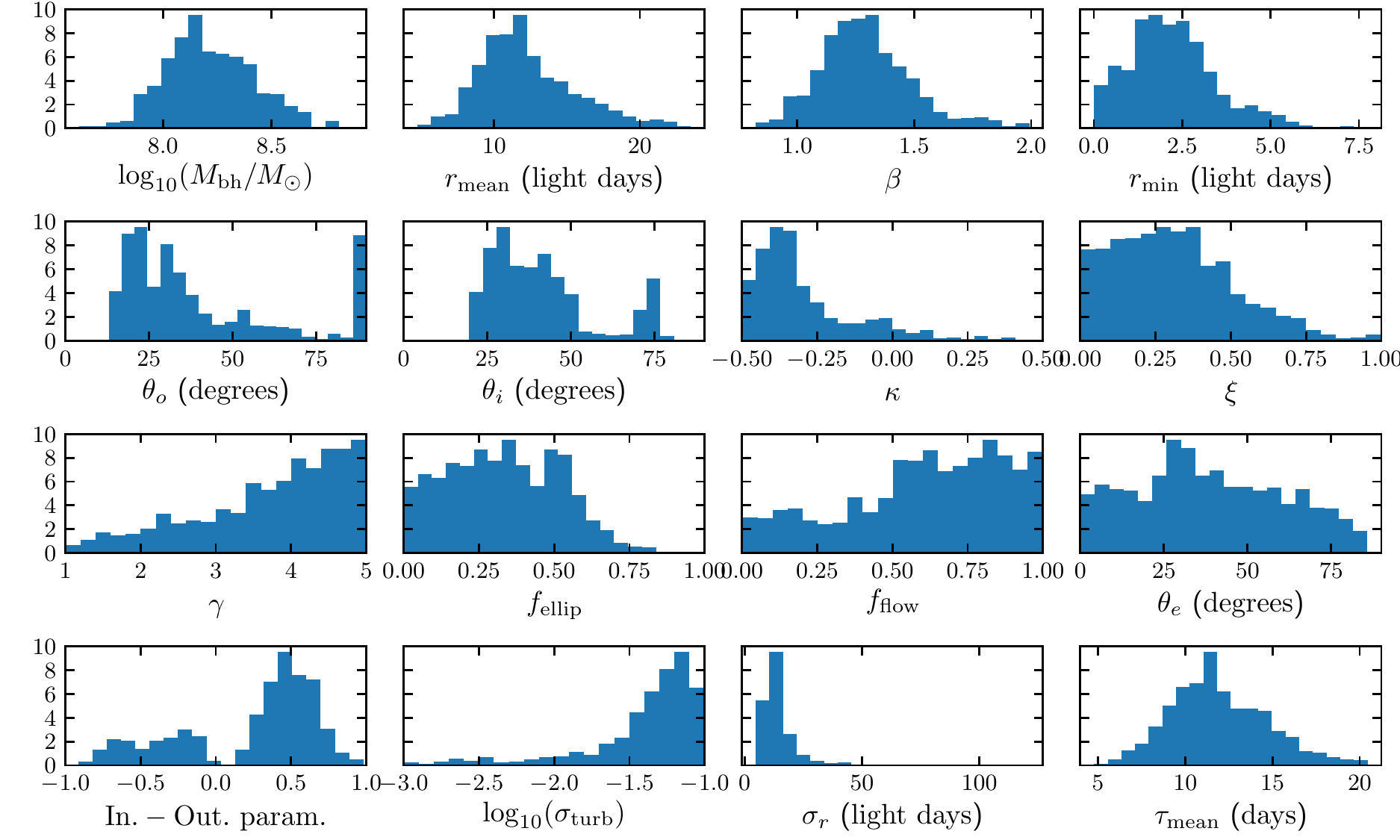}
    \caption{Posterior distributions of select model parameters for the QSO model.}
    \label{fig:posterior_caramelqso}
\end{figure*}

The CARAMEL BLR model is able to reproduce both the H$\alpha$ line profile shape as well as the integrated H$\alpha$ flux light curve for the QSO model (Figure~\ref{fig:display_caramelqso}). Below, we discuss the modelling results, giving median values and 68\% confidence intervals for the key model parameters. In cases where the posterior PDF is one-sided, upper and lower 68\% confidence limits are given instead. The full posterior PDFs are provided in Figure \ref{fig:posterior_caramelqso}.

The CARAMEL modelling results for the QSO model show an H$\alpha$-emitting BLR that is steeper than exponential with a shape parameter $\beta = $~\cqsocombbeta\ for the Gamma distribution.
The distribution is shifted from the origin by $r_{\rm min} = $~\cqsocombrmin\ light days.
The mean and median radii are $r_{\rm mean} = $~\cqsocombrmean\ and $r_{\rm median} = $~\cqsocombrmedian\ light days, and the radial thickness of the distribution is $\sigma_r = $~\cqsocombsigmar\ light days.
The mean and median lags are $\tau_{\rm mean} = $~\cqsocombtaumean\ and $\tau_{\rm median} = $~\cqsocombtaumedian\ days. The full posterior PDFs show multiple solutions for the opening angle and inclination angle, so a spherical BLR is not completely ruled out. Taking the median value and $68\%$ confidence intervals, we find an H$\alpha$-emitting BLR that is a thick disc, with $\theta_o = $~\cqsocombthetao\ degrees. The disc is inclined relative to the observer's line of sight by an inclination angle $\theta_i = $~\cqsocombthetai\ degrees. The 2d posterior PDF for $\theta_i$ vs. $\theta_o$ shows that the models in which $\theta_{i} \approx 75$ degrees are the same models for which $\theta_{o} \rightarrow 90$ degrees.

The emission comes preferentially from the far side of the BLR ($\kappa = $~\cqsocombkappa), which is what one would expect if the BLR clouds emit preferentially back towards the ionizing source. Models with an opaque disc mid-plane are slightly preferred over those with a transparent mid-plane ($\xi = $~\cqsocombxi).
Finally, the results show that the emission comes preferentially from the faces of the disc ($\gamma \geq 3.4$), making the geometry closer to a cone than a uniformly distributed thick disc.

Dynamically, models are preferred in which there is a mixture of gas on near-circular elliptical orbits and gas in inflowing or outflowing trajectories ($f_{\rm ellip} = $~\cqsocombfellip).
There is a slight preference for outflowing gas over inflowing gas ($f_{\rm flow} = $~\cqsocombfflow) for the gas that is not on near-circular elliptical orbits. We note that since $\theta_e = $~\cqsocombthetae\ degrees, the radial and tangential velocities are drawn from a distribution that may be rotated a non-negligible angle towards the circular velocity, resulting in fewer of the BLR particles with truly unbound trajectories. To summarize the total amount of inflowing or outflowing gas, we calculate an additional parameter:
\begin{align}
{\rm In. - Out.} = {\rm sgn}(f_{\rm flow} - 0.5) \times (1 - f_{\rm ellip}) \times \cos\theta_e,
\end{align}
where ${\rm sgn}$ is the sign function.
This parameter is constructed such that a BLR with pure radial outflow (inflow) will have ${\rm In. - Out.} = 1~(-1)$, and a BLR with no preference for either solution will have ${\rm In. - Out.} = 0$.
The posterior distribution for this parameter shows solutions for both inflow and outflow, although those with outflowing gas are preferred. Finally, we find that macro-turbulence may be important to the dynamics, with $\sigma_{\rm turb} = $~\cqsocombsigmaturb * $v_{\rm circ}$. The black hole mass inferred for this model is $\log_{10}(M_{\rm BH}/M_\odot) = $~\cqsocomblogmbh.

\begin{figure}
	\includegraphics[width=\columnwidth]{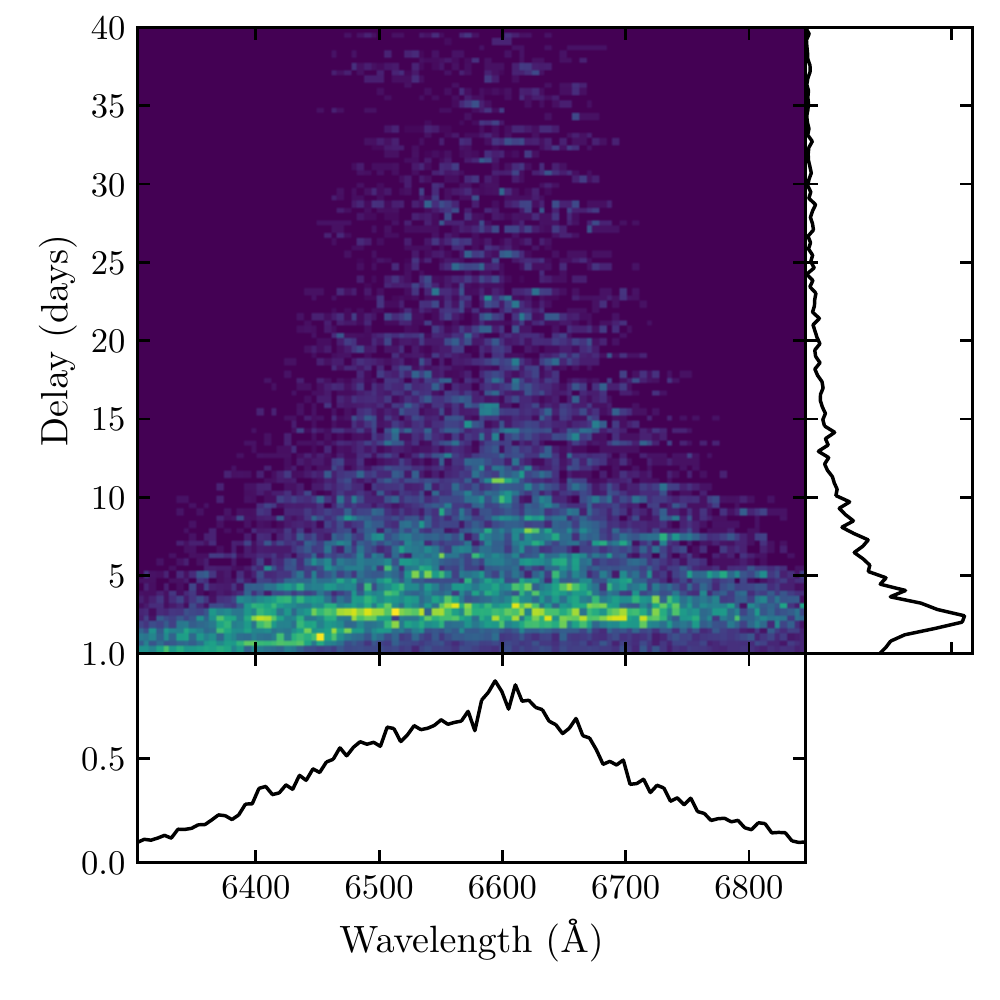}
    \caption{Velocity-resolved transfer function for the QSO model, chosen to be representative of the full posterior sample. The right-hand panel shows the velocity-integrated transfer function and the bottom panel shows the time-averaged line profile.}
    \label{fig:transfer_caramelqso}
\end{figure}

\begin{figure}
	\includegraphics[width=0.9\columnwidth]{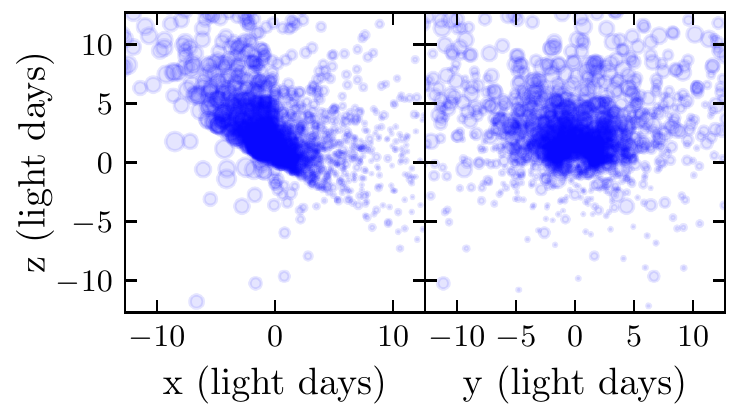}
    \caption{Geometric model of the broad line region that was used to create the transfer function in Figure~\ref{fig:transfer_caramelqso}. Each circle represents one of the particles in the model, and the size of the circle is proportional to the relative strength of emission from the particle, as determined by Equation \ref{eq:caramel_weight}. The observer is situated along the positive $x$-axis.}
    \label{fig:geo_caramelqso}
\end{figure}

\subsection{Unblinding: Comparison to Ground Truth}
\label{sec:results:models}

\subsubsection{\memecho\ vs ground truth}
\label{sec:results:models:memecho}

The primary output of the \memecho\ analysis is the recovered 2-D response function, shown in Figure~\ref{fig:twodqsomap}. This can be directly compared to the true (input) response function in Figure~\ref{fig:method:response_qso}.

In our view, the performance of \memecho\ in recovering the input velocity-delay map is quite impressive. To a good approximation, the recovered map is a smoothed version of the input map, exactly as one might hope and expect. However, the true peak in the overall delay distribution lies at $\simeq 4$~days, whereas the peak in the recovered distribution is estimated to be $\simeq 6$~days. This is almost certainly associated with the inevitable smoothing associated with regularization and exacerbated by the skewness of the delay distribution. The \memecho\ velocity-delay map correctly reproduces the shape of the virial envelope in the input map, as well as the weakness of the response near line centre. The difference in the responses of the two line wings -- with the red wing exhibiting a stronger response than the blue wing -- is also captured correctly in the \memecho\ map.

Turning to the physical interpretation of the \memecho\ results provided by K.H., he notes that the dominant structure in the recovered velocity-delay map can be explained by a BLR that consists mainly of gas extending from 5 to 15~light days in Keplerian rotation around a $M_{\rm BH} \simeq 10^8$~M$_{\odot}$ black hole, viewed from an inclination of $i \simeq 45^{\circ}$. Comparing this to Figures~\ref{fig:geometry-emissivity} and ~\ref{fig:geometry-responsivity}, as well as to the numbers in Table~\ref{table:params}, we see that this interpretation does capture the basic shape and kinematics of the line-forming region. More specifically, in our model, the emission line (and its response) are formed primarily in the dense, rotation-dominated base of the outflow. The geometry of and kinematics in this region are indeed similar to those of an annulus in Keplerian rotation between $r_{\rm min}$ and $r_{\rm max}$, and our adopted viewing angle is $i = 40^{\circ}$, similar to that inferred from the \memecho results.

The main discrepancy between the physical interpretation of the \memecho\ results and the input model concerns the physical scale of the line-forming region. Figures~\ref{fig:geometry-emissivity} and ~\ref{fig:geometry-responsivity} show that -- in line with $r_{\rm min}$ and $r_{\rm max}$ in Table~\ref{table:params} (3.29-6.602 light days)-- the actual radius of the line-forming ``annulus" in our model is roughly 2/3 of that inferred from the \memecho\ reconstruction (i.e. $\simeq 7$~light-days~$=1.8\times 10^{16}$~cm). This is exactly in line with the factor of $\simeq 50\%$ difference between the true and inferred mean delays, and is therefore also presumably caused by the effective smoothing of the response function during the maximum entropy inversion.

Given that the BLR radius is overestimated, we might have expected the black hole mass to be overestimated also (since the virial estimator scales as $M_{\rm BH} \propto v^2 R$). However, the estimate obtained from the \memecho\ reconstruction is $M_{\rm BH} \simeq 10^{8}$~M$_{\odot}$, whereas the black hole mass in our (rescaled) model is $M_{\rm BH} \simeq 2\times 10^8$~M$_{\odot}$. Based on Figure~\ref{fig:twodqsofit}, the main reason for this difference appears to be that the \memecho\  estimate is derived from the {\em outer envelope} of the brightest parts of the 2-D response function. This outer envelope lies at velocities that are higher than typical for the bulk of the line-forming region. This, coupled with the slightly overestimated inclination, biases the \memecho\ black hole mass estimate towards higher values and more than compensates for the effect of the overestimated BLR radius. In any case, agreement to within a factor of $\simeq 2$ is in line with the accuracy expected for this qualitative assessment.

We finally note that neither the \memecho\ velocity-delay map itself, nor its interpretation by an expert, point towards a rotating {\em outflow} as the source of the variable emission line. As noted above, given that the line formation in our disc wind model takes place primarily within the dense, rotation-dominated base of the outflow, this should not come as a surprise. It is nevertheless important to keep this in mind when interpreting observational data: physically motivated BLR models can have complex geometries and kinematics that may not be easy to discern even from 2-D response functions. Comparisons with toy models -- e.g. Hubble inflows/outflows, pure Keplerian discs -- may still provide useful insights in these cases. However, it is crucial to remember that we are only studying those parts of the BLR that dominate the responsivity-weighted line emission. Even if the inferred geometry and kinematics for these regions are broadly correct, they may not reflect the overall geometry and kinematics of the flow that constitutes the BLR.

\subsubsection{CARAMEL vs ground truth}
\label{sec:results:models:caramel}

The primary output of the CARAMEL analysis is the set of parameter distributions shown in Figure~\ref{fig:posterior_caramelqso} and discussed in Section~\ref{sec:results:caramel}. These parameters define the properties of the cloud population used by CARAMEL to fit the simulated data. The overall geometry of this population is shown in Figure~\ref{fig:geo_caramelqso}, which can be compared to our raw and responsivity-weighted emissivity maps (Figure~\ref{fig:geometry-emissivity}~and~\ref{fig:geometry-responsivity}).

Even though a spherical BLR was not completely ruled out by the CARAMEL modelling, the preferred geometry was a strongly flared disc (opening angle $\theta_{o} = 32^{+36}_{-12}$~degrees viewed at an inclination of $i \simeq 40^{\circ}$. The inferred inclination is in excellent agreement with the true value, and a flared disc is a reasonable description of the line-forming region in our biconical disc wind model. Indeed, line emission in the CARAMEL models is produced preferentially near the face of the disc, in line with a conical geometry. In the model, the inner part of the wind cone lies at an angle of $90^{\circ}-\theta_{\rm min} = 20^{\circ}$ from the disc surface, which is smaller than, but still consistent with, the inferred opening angle.

The velocity-integrated median delay obtained by the CARAMEL analysis is $\tau_{\rm median} = 6.6^{+1.9}_{-1.5}$~days. This agrees well with the actual median delay $\tau_{\rm median} \simeq 6$~day. In line with this, the characteristic scale of the line-forming region is also correctly recovered, $r_{\rm median} \simeq 7$~light days, in good agreement with the approximate radius of the line-emitting annulus in our model (see Figures~\ref{fig:geometry-emissivity} and ~\ref{fig:geometry-responsivity}). CARAMEL also correctly finds that the line emission comes preferentially from the far side of the BLR, and that the mid-plane of the disc is opaque.

Turning to the kinematics of the BLR, the picture is less clear. CARAMEL correctly finds that a significant part of the BLR material is on near-circular Keplerian orbits and also that an additional velocity field is required. However, it cannot decisively distinguish between inflow and outflow kinematics, even though it does (correctly) favour a net outflow of material. CARAMEL also finds marginal evidence for a significant macro-turbulent velocity, which we suspect is an artefact of its kinematic parameterization being unable to faithfully describe the ``true" BLR kinematics. The black hole mass of $M_{\rm BH} \simeq 2 \times 10^8$~M$_{\odot}$ is correctly recovered by CARAMEL.

Perhaps the most surprising and concerning aspect of the CARAMEL analysis is the velocity-delay map constructed from a model drawn at random from the posterior distribution  (Figure~\ref{fig:transfer_caramelqso}). Some difference would be expected as CARAMEL assumes that emission is \emph{not} linearised around a mean line luminosity, however this velocity-delay map looks completely different from both our input response function (Figure~\ref{fig:method:response_qso}) and the response function recovered by \memecho\ (Figure~\ref{fig:twodqsomap}). For example, it does not recover the double-peaked nature of the response, i.e. the suppressed response near line centre. It also shows no bright emission from the virial envelope associated with any particular annulus, just a smooth distribution across the entire width of the envelope at long delays, and a bright, diagonal ``line" at short delays (with a blue-leads-red signature).

We have checked whether the particular model shown in Figure~\ref{fig:transfer_caramelqso} was just an ``unlucky" draw from the posterior distribution, i.e. that it is not representative. We find that other models drawn from the posterior distribution can exhibit (weakly) double-peaked mean line profiles, but the velocity-delay maps always tend to be quite similar to Figure~\ref{fig:transfer_caramelqso} (and hence dissimilar to the input response function).

\begin{figure}
	\includegraphics[width=\columnwidth]{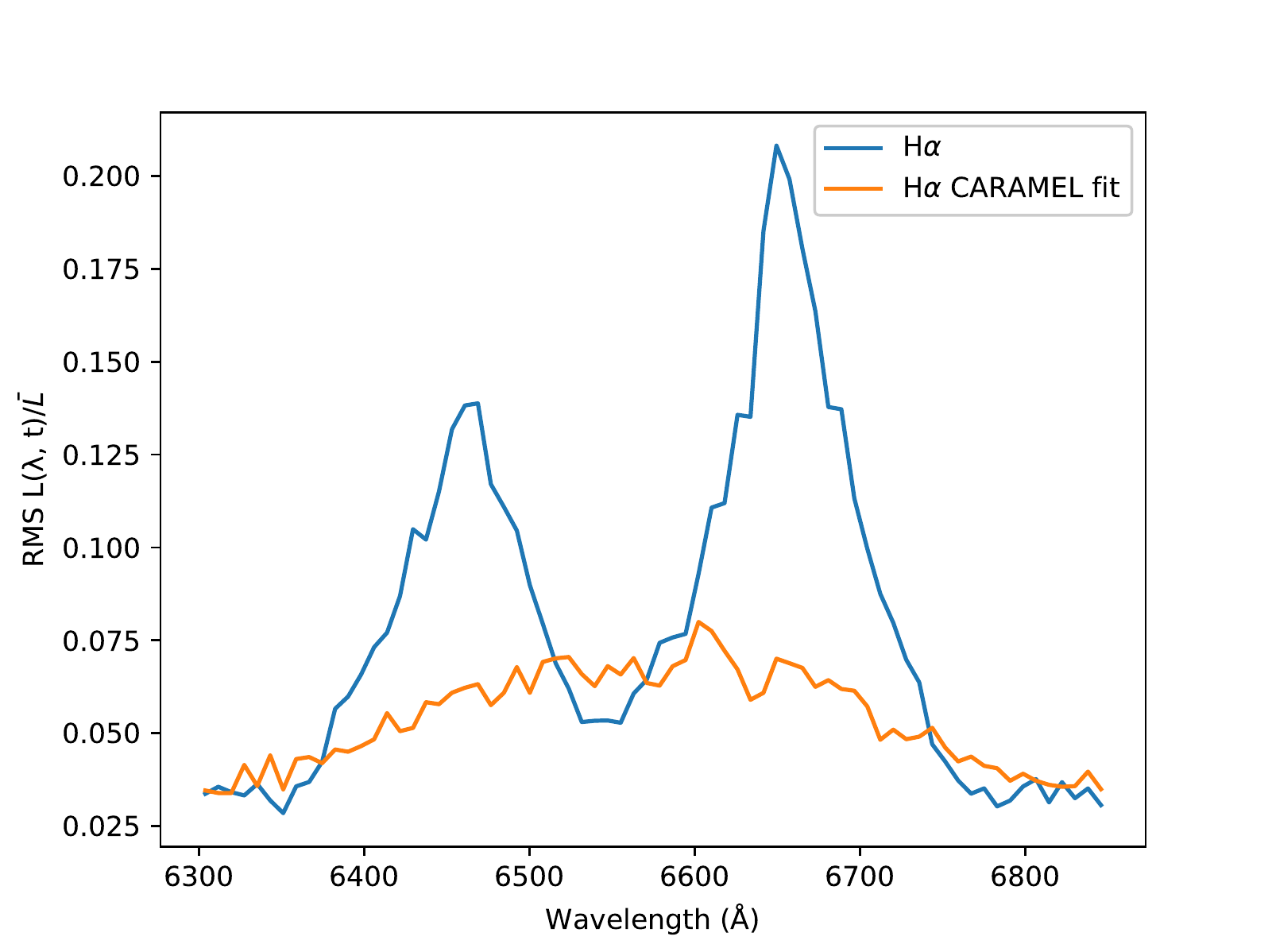}
	\caption{RMS residuals for the noisy output time series of spectra and CARAMEL fit to it.}
    \label{fig:discussion:rms}
\end{figure}

\begin{figure}
	\includegraphics[width=\columnwidth]{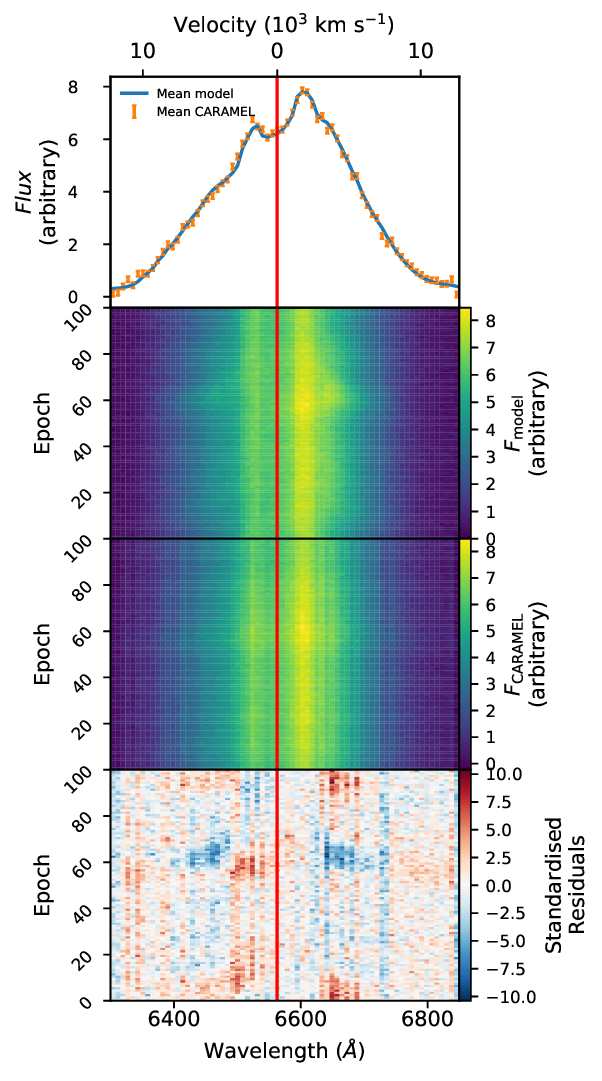}
    \caption{CARAMEL model fits to the QSO model H$\alpha$ line profile, compared to the model line profiles. Red lines indicate line centres.
	Panel $1$: The mean model and CARAMEL line profiles across all epochs.
	Panel $2$: The model emission-line profile for each epoch.
	Panel $3$: The CARAMEL fit emission-line profile for each epoch.
	Panel $4$: The standardised model - fit residuals for the emission-line profile for each epoch.
	}
    \label{fig:discussion:caramel_residuals}
\end{figure}

Given that the velocity-delay map is just a by-product of the CARAMEL analysis, it is important to test whether this discrepancy is associated purely with the construction of the response function from the CARAMEL models (rather than the models themselves). What is actually being fit by CARAMEL is the spectroscopic time series itself. We have therefore also constructed the RMS line profile directly from the input time series and also from the CARAMEL model fit to this time-series. These RMS line profiles are shown in Figure~\ref{fig:discussion:rms}. It is immediately apparent that there does seem to be a fundamental problem with the CARAMEL model fits: the RMS profile constructed from the CARAMEL model has a completely different shape than that constructed from the input data. Most importantly, the CARAMEL RMS profile is single-peaked, whereas the input disc wind model produces a clearly double-peaked RMS profile (as we would expect, given its rotation-dominated kinematics).  The standardized model - fit residuals for the time series of spectra go some way further illustrate this deviation (Figure~\ref{fig:discussion:caramel_residuals}); it is apparent that whilst CARAMEL can capture the variation around the line peak well, variations of the regions on either side of the line peak are poorly-captured. CARAMEL simultaneously under- and over-predicts, unable to capture the asymmetry to the line response.

We currently have no simple explanation for this behaviour. It is perplexing that a model that produces a response function and RMS profile that appear to be inconsistent with ground truth should nevertheless be able to match the individual spectra in our time series (as suggested by Figure~\ref{fig:display_caramelqso}). The discrepancy is potentially due to CARAMEL's use of \emph{linear} line response to the continuum, rather than \emph{linearised} response around the mean flux as assumed by \memecho\ . This would explain why CARAMEL can fit the \emph{mean} line profile relatively well, but fail to capture variations around that mean.

\section{Conclusions}
\label{sec:conclusions}

We have tested the ability of the two main inversion techniques used in AGN reverberation mapping to recover two physically-motivated response functions from a simulated time-series of spectra. The two reverberation mapping codes we tested were \memecho\ and CARAMEL which represent two different classes of inversion techniques. \memecho\ simply aims to recover the 2-D response function, with the physical interpretation of the results being left to the expert user. CARAMEL carries out forward modelling of the spectroscopic time-series, using a simple, but flexible, description of the BLR as a population of orbiting clouds with known geometric and kinematic properties. All tests were carried out as blind trials, i.e. the \memecho\ and CARAMEL modelling teams were only provided the simulated time series.

The benchmark BLR models used in our test describe a rotating, biconical accretion disc wind. The simulated spectroscopic time series were generated from a self-consistent ionization and radiative transfer simulation that follows the H$\alpha$ line formation process within the outflow. Two different sets of model parameters were used as input, which were roughly designed to represent Seyfert galaxies and QSOs. In both models, the H$\alpha$ line-forming region lies primarily in the dense base of the wind, where the kinematics are rotation-dominated.

Neither the maximum-entropy technique of \memecho\ nor Markov-Chain Monte-Carlo forward modelling technique of CARAMEL were able to successfully recover the Seyfert response function, due to the significant negative responsivity in large parts of the velocity-delay space of this model. However, both methods fail ``gracefully'', in the sense of not generating spurious results.

In the case of the QSO model, the velocity-delay map recovered by \memecho\ was a good match to the input 2-D response function, after accounting for the inevitable smoothing associated with the inversion process. The expert interpretation of the map also correctly captured the annular geometry and rotation-dominated kinematics of the line-forming region. In addition, the estimated observer orientation and black hole mass were in reasonable agreement with those in the input model. The characteristic size of the BLR was overestimated by roughly 50\%, however.

CARAMEL also captured the overall geometry of the line-forming region, describing it as a flared, inclined disc with the correct size and orientation. The importance of rotation to the kinematics was also recovered, but while an additional kinematic component was required by the modelling, CARAMEL was unable to reliably distinguish between inflow and outflow velocity fields for this component. Nevertheless, the black hole mass was correctly estimated by CARAMEL.

The most surprising and concerning result of the CARAMEL analysis is that the velocity-delay map it recovers is strongly inconsistent with the true 2-D response function. In line with this, the RMS profile of the CARAMEL fits to the spectroscopic time series is also inconsistent with that of the input time series. We currently have no explanation for these discrepancies. They are difficult to understand in light of the apparently successful fits CARAMEL achieves to the individual spectra.

Overall, we consider the results of these tests to be quite positive. Even though neither model was able to deal with the Seyfert model, with its net negative response, neither generated misleading results in this case. In the case of the QSO model, both methods broadly recovered the correct geometry of the line-forming region, as well as its dominant kinematics. However, neither method was able to capture that the input model described a disc wind. This should not come as a surprise, given that the rotation dominates the kinematics in the line-forming region. It is nevertheless critical to keep this lesson in mind when interpreting observational data sets: even correctly recovered and interpreted response functions can only tell us about the conditions in the (responsive parts of the) line-forming region. This region can be dominated by rotation, for example, even if this part of the BLR is just the inner part of a larger-scale outflow.

The results do raise the concern that real time-series that exhibit negative responses (as suggested by \citet{Pei2017}) cannot be reliably analysed by current techniques. This suggests that the inversion method of regularized linear inversion \citep{Krolik1995}, which has the capability to handle negative responses, warrants revisiting. Promisingly, the \memecho\ team are also exploring modifications to the code that would enable it to handle such time series. Similarly, the CARAMEL team are currently implementing photoionization physics into their model, which has the potential to introduce a spatially-dependent responsivity into the model, allowing it to handle negative responses.

This work only covers the analysis of two possible models with a single continuum, and whilst the simulated observing campaigns satisfy the criteria in \citet{Horne2004}, we believe simulating campaigns with a more diverse range of models and (in particular) continuum variation profiles would allow us to more closely define the limits of existing deconvolution techniques. This could potentially give this method a role in the planning of observational campaigns, allowing them to place bounds on their capabilities.

\section*{Acknowledgements}
Most figures in this paper were prepared using \emph{PGPLOT} \citep{Pearson2017}, \emph{MatPlotLib} \citep{Hunter2007} and \emph{Dia Diagram Editor} \citep{Oualline2018}. SWM acknowledges the University of Southampton`s Institute for Complex Systems Simulation and the Engineering and Physical Sciences Research Council for the PhD student that funded his research.
CK and NSH acknowledge support by the Science and Technology Facilities Council grant ST/M001326/1. JHM is supported by the Science and Technology Facilities Council under grant ST/N000919/1. KH acknowledges support from STFC grant ST/R000824/1.

\section{Appendix}
As discussed in Section~\ref{sec:method:tss}, in order to test the ability of deconvolution codes to our response functions it is necessary to provide them with a synthetic observing campaign, in the form of a time series of spectra.

\begin{figure*}
	\includegraphics[width=\textwidth]{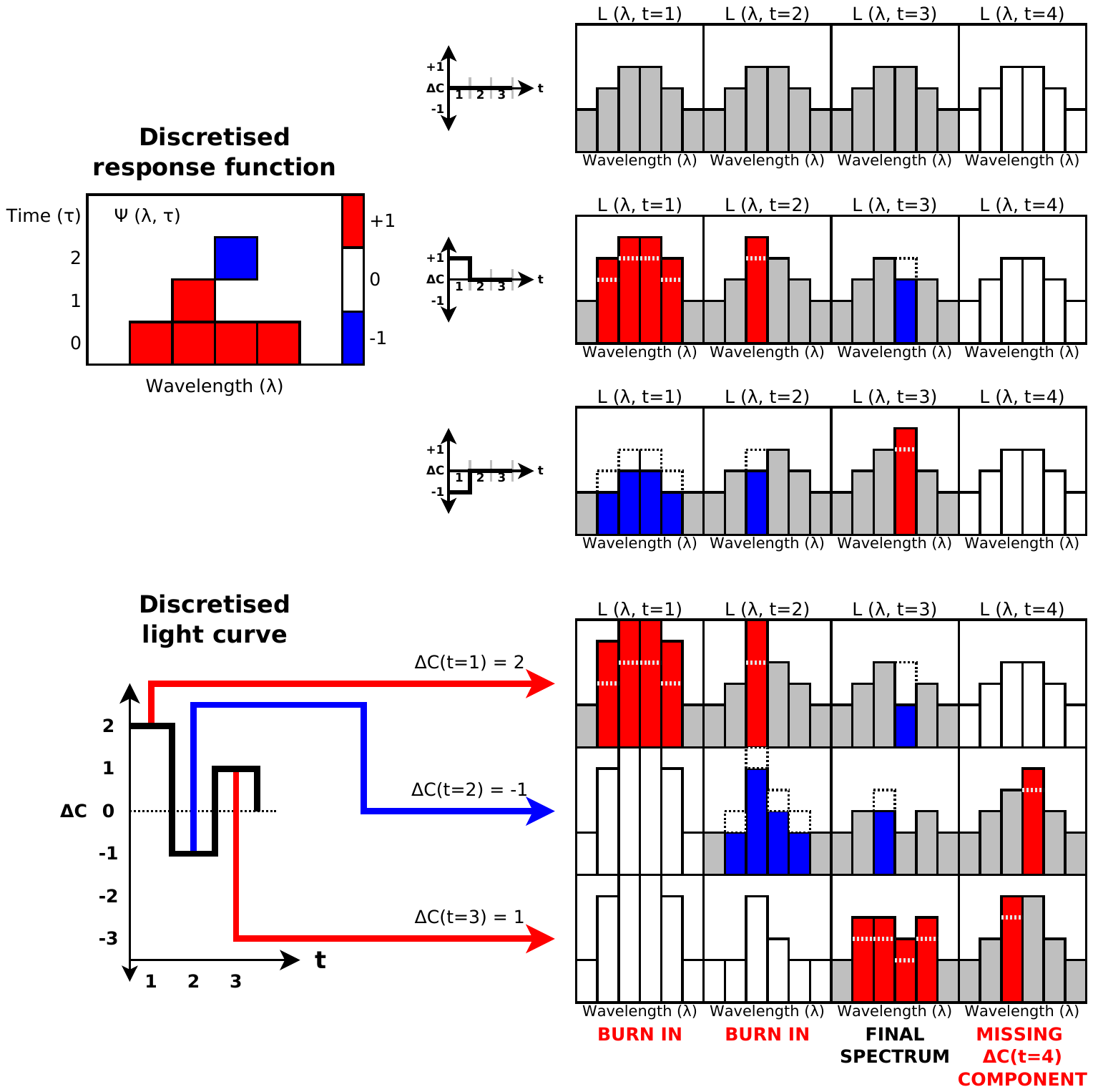}
	\caption{Diagram showing how a discretised toy response function $\Psi_R(\lambda, \tau)$ (top left) applies changes in continuum luminosity $\Delta C$ to a base spectrum to form a series of spectra at later. Top right 3 panels illustrate how a change in luminosity in a single time step $t$ propagates out over a range of times $t+\tau$. Within a spectrum, columns in \textbf{\textcolor{blue}{blue}} indicate a wavelength bin whose flux has been \emph{decreased} by a negative response at this $\tau$ and $\lambda$, columns in \textbf{\textcolor{red}{red}} a wavelength bin whose flux has been \emph{increased}, and columns in grey a wavelength bin whose flux is unchanged. Columns in \textbf{white} are those outside of the $t+\tau$ range that the response function applies to, and so that were \emph{ineligible} to be changed this time-step. The lower panels illustrate how a discretised light curve (bottom left) can be used to produce a final spectrum (bottom right). Only spectra for time-steps $t > t_0+\tau_\textrm{\rm max}$ and $t < t_{S,\textrm{\rm max}}$ are \emph{fully-constructed} spectra.}
    \label{fig:method:detail}
\end{figure*}

The code for producing these synthetic spectroscopic time-series is written in \emph{Python} using the \emph{Numpy} \citep{Oliphant2015} and \emph{Astropy} \citep{TheAstropyCollaboration2018} modules. We initialise each spectrum in the time series to a base spectrum taken from an assumed mean luminosity model. Then, we create a grid in time between the initial and last continuum measurement, with spacing $\Delta t$ equivalent to the smallest bin in our discretised response function, $\Psi_{RD}(v,t)$. For each step in this grid, $t_i=t_0+i \, \Delta t$, we perform simple linear interpolation to arrive at the value of the driving continuum, $C$, and thus the deviation from the mean luminosity, $\Delta C(t_i)$. Then, starting at the initial time-step, we apply the contribution from the change in luminosity, $\Delta C$, to the spectrum at each later time-step, $t_i + \tau$, as $\Psi_{RD}(v, \tau) \, \Delta C(t_i)$. Continuing this process for every time-step yields the emission line at time $t$ as
\begin{equation}
	L(v,t) = L_0(v) + \sum_{t_i} \Psi_{RD}(v, t-t_i) \, \Delta C (t_i) \, \Delta t,
	\label{eqn:method:timesteps}
\end{equation}
which is the mirrored, discrete equivalent to Equation~\ref{eqn:method:timesteps}. A diagram of this process for a toy transfer function is shown in Figure~\ref{fig:method:detail}. The physically-motivated design behind this implementation (seeking to mimic the actual process giving rise to the response function) makes it relatively intuitive and offers the scope for further development. In particular, it is straightforward to allow for \emph{continuum-dependent response functions} in this framework, though that is not within the scope of the present work.

CARAMEL works on continuum-subtracted spectra. When generating time-series for CARAMEL, we therefore set $L_0$ to the continuum-subtracted base spectrum from the mean luminosity model. A simple linear fit is used to subtract the continuum under the line. By contrast, \memecho\ is designed to work on a sequence of spectra that include the (variable) local continuum. Denoting line plus continuum spectra as $\mathcal{L} = C + L$, the input we provide to \memecho\ can be written as
\begin{equation}
	\mathcal{L}(v,t) = \mathcal{L}_0(v) + \Delta C(t) + \sum_{t_i} \, \Psi_{RD}(v, t-t_i) \, \Delta C (t_i) \, \Delta t.
	\label{eqn:method:timesteps_memecho}
\end{equation}
Here, $\mathcal{L}_0(v) = C_0 + L_0(v)$, is just the base spectrum provided by {\sc Python} for our mean luminosity model.

As previously discussed, our driving continuum $C(t)$ is a rescaled version of the 1158~\AA\ light-curve of NGC~5548 from \citet{DeRosa2015a}. It is rescaled to match the mean luminosity for each model, and the range of variation is reduced to $\pm50\%$ about this mean value. As the response function reprocesses this light-curve to produce the time series of spectra, we cannot produce meaningful line profiles for dates before $t_{\rm start}+t_{\rm\Psi max}$, as insufficient continuum measurements are available to reprocess. This affects not only our ability to simulate spectra, but any attempt to invert observational spectroscopic time-series. Here, we deal with this by means of a ``burn-in period", as illustrated in Figure~\ref{fig:method:detail}.

We finally add simulated observational errors to the spectra for each time series. We adopt a constant absolute error in each spectral bin of each spectrum in the simulated time series. For the QSO model, the actual level of added noise is set so that the error on the integrated line flux measured from a single spectrum is, on average, 2\% of the peak-to-peak variation in the integrated line flux. Since this peak-to-peak variation is much smaller for our Seyfert model -- as a consequence of its globally small and negative response (c.f. Figure~\ref{fig:method:response_sey}), this method would produce unrealistically small errors in this case. We therefore instead set the noise in our Seyfert time series such that, on average, the signal-to-noise near the peak of the line in a single spectrum for the Seyfert model is the same as for the QSO model. We also assume that the noise is drawn from a Gaussian distribution and that all of the uncertainties are uncorrelated.

The output of our method is two separate time series of spectra, one continuum-subtracted, one not. Both consist of 101 line profiles, simultaneous with the last 101 of the 176 continuum measurements, with noise added.





\bibliographystyle{mnras}
\bibliography{mybib} 



\bsp	
\label{lastpage}
\end{document}